\documentstyle[12pt,epsf]{article}
\def\hybrid{\topmargin -20pt    \oddsidemargin 0pt
        \headheight 0pt \headsep 0pt
        \textwidth 6.5in        
        \textheight 9in         
        \marginparwidth .875in
        \parskip 5pt plus 1pt   \jot = 1.5ex}

\hybrid
\makeatletter
\@addtoreset{equation}{section}
\makeatother

\newcommand{\cC}{{\cal C}}

\newcommand{\cK}{{\cal K}}

\newcommand{\cM}{{\cal M}}
\newcommand{\cN}{{\cal N}}

\newcommand{\cR}{{\cal R}}

\newcommand{\qt}{\frac14}
\newcommand{\bea}{\begin{eqnarray}}
\newcommand{\eea}{\end{eqnarray}}
\newcommand{\be}{\begin{equation}}
\newcommand{\ee}{\end{equation}}
\newcommand{\bt}{\begin{tabular}}
\newcommand{\et}{\end{tabular}}
\newcommand{\ba}{\begin{array}}
\newcommand{\ea}{\end{array}}

\newcommand{\bmat}{\left(\begin{array}}
\newcommand{\emat}{\end{array}\right)}
\newcommand{\vev}[1]{\langle #1 \rangle}

\newcommand{\Tr}{\mathop{\rm Tr}}

\newcommand{\rank}{\mathop{\rm rank}}
\newcommand{\diag}{{\rm diag}\,}
\newcommand{\R}{{\rm Re}}

\def\beq{\begin{equation}}
\def\eeq{\end{equation}}
\def\beqa{\begin{eqnarray}}
\def\eeqa{\end{eqnarray}}

\def\NPB#1#2#3{Nucl.\ Phys.\ B{#1} (19#2) #3}
\def\PLB#1#2#3{Phys.\ Lett.\ B{#1} (19#2) #3}

\def\JGP#1#2#3{J.\ Geom.\ Phys.\ {#1} (19#2) #3}

\def \N#1{$\cN=#1$}
\def \D#1{$D=#1$}

\def\yzero{\smash{\hbox{$y\kern-4pt\raise1pt\hbox{${}^\circ$}$}}}

\def\eps{\epsilon}

\def\s2{\frac{1}{\sqrt2}}

\def\bZ{{\mathbf Z}}
\def\ZNM{$\IZ_N\times\IZ_M$}
\def\half{\frac12}

\def\IB{\relax{\rm I\kern-.18em B}}
\def\ID{\relax{\rm I\kern-.18em D}}
\def\IE{\relax{\rm I\kern-.18em E}}
\def\IF{\relax{\rm I\kern-.18em F}}
\def\IH{\relax{\rm I\kern-.18em H}}
\def\II{\relax{\rm I\kern-.18em I}}
\def\IK{\relax{\rm I\kern-.18em K}}
\def\IL{\relax{\rm I\kern-.18em L}}
\def\IM{\relax{\rm I\kern-.18em M}}
\def\IN{\relax{\rm I\kern-.18em N}}
\def\IP{\relax{\rm I\kern-.18em P}}
\def\IR{\relax{\rm I\kern-.18em R}}
\def\IT{\relax{\rm I\kern-.42em T}}
\def\IZ{\relax{\hbox{\raisebox{.38ex}
    {\scriptsize\bfseries\slshape /}\kern-.40em\_\kern-.28em\rm Z}}}
\def\Iz{\relax{\hbox{\raisebox{.38ex}
    {\tiny\bfseries\slshape /}\kern-.25em\raisebox{.65ex}
    {\tiny\bfseries\slshape /}\kern-.43em\_\kern-.26em\rm Z}}}
\def\inbar{\vrule height1.5ex width.8pt depth-0.2pt}
\def\inbarhi{\vrule height1.55ex width.5pt depth-.85ex}
\def\inbarlo{\vrule height.8ex width.5pt depth0ex}
\def\IC{\relax{\rm C\kern-.48em \inbar\kern.48em}}
\def\IO{\relax{\rm O\kern-.56em \inbar\kern.56em}}
\def\IQ{\relax{\rm Q\kern-.56em \inbar\kern.56em}}
\def\IS{\relax{\rm S\kern-.37em \inbarhi\kern.08em\inbarlo\kern.29em}}

\def \one{\relax{\rm 1\kern-.26em I}}

\def\Dsl{\,\raise.15ex\hbox{/}\mkern-13.5mu D} 

\newcommand{\drawsquare}[2]{\hbox{%
\rule{#2pt}{#1pt}\hskip-#2pt
\rule{#1pt}{#2pt}\hskip-#1pt
\rule[#1pt]{#1pt}{#2pt}}\rule[#1pt]{#2pt}{#2pt}\hskip-#2pt
\rule{#2pt}{#1pt}}

\newcommand{\fund}{\raisebox{-.5pt}{\drawsquare{6.5}{0.4}}}
\newcommand{\Ysymm}{\raisebox{-.5pt}{\drawsquare{6.5}{0.4}}\hskip-0.4pt%
        \raisebox{-.5pt}{\drawsquare{6.5}{0.4}}}
\newcommand{\Yasymm}{\raisebox{-3.5pt}{\drawsquare{6.5}{0.4}}\hskip-6.9pt%
        \raisebox{3pt}{\drawsquare{6.5}{0.4}}}
\newcommand{\antifund}{\overline{\fund}}
\newcommand{\bYasymm}{\overline{\Yasymm}}
\newcommand{\bYsymm}{\overline{\Ysymm}}

\def \nonab{non\discretionary{-}{}{-}Abel\-ian\ }
\def \nonvan{non\discretionary{-}{}{-}van\-ish\-ing\ }
\def \noneq{non\discretionary{-}{}{-}equi\-va\-lent\ }
\def \nonze{non\discretionary{-}{}{-}ze\-ro\ }
\def \nontriv{non\discretionary{-}{}{-}triv\-i\-al\ }

\begin{document}

\pagestyle{empty}
\renewcommand{\thefootnote}{\fnsymbol{footnote}}
\rightline{FTUAM-00/06, IFT-UAM/CSIC-00-08}
\rightline{\tt hep-th/0002103}
\vspace{0.5cm}
\begin{center}
\LARGE{\bf Orientifolds with discrete torsion \\[20mm]}

\large{M.~Klein, R.~Rabad\'an\\[5mm]}

\small{Departamento de F\'{\i}sica Te\'orica C-XI
       and Instituto de F\'{\i}sica Te\'orica  C-XVI,\\[-0.3em]
       Universidad Aut\'onoma de Madrid,
       Cantoblanco, 28049 Madrid, Spain.\footnote{E-mail: 
       matthias.klein@uam.es, rabadan@delta.ft.uam.es}\\[20mm]}

\small{\bf Abstract} \\[7mm]
\end{center}

\begin{center}
\begin{minipage}[h]{14.0cm}

{\small
We show how discrete torsion can be implemented in \D4, \N1 type IIB 
orientifolds. Some consistency conditions are found from the closed string
and open string spectrum and from tadpole cancellation. Only real values 
of the discrete torsion parameter are allowed, i.e.\ $\eps=\pm1$. Orientifold 
models are related to real projective representations. In a similar way as 
complex projective representations are classified by 
$H^2(\Gamma,\IC^*)=H^2(\Gamma,U(1))$, real projective representations are 
characterized by $H^2(\Gamma,\IR^*)=H^2(\Gamma,\IZ_2)$. Four different types 
of orientifold constructions are possible. We classify these models and give 
the spectrum and the tadpole cancellation conditions explicitely.
}
\end{minipage}
\end{center}
\newpage

\setcounter{page}{1}
\pagestyle{plain}
\renewcommand{\thefootnote}{\arabic{footnote}}
\setcounter{footnote}{0}


\section{Introduction}

On analyzing the effect of a \nonvan B-field on the modular invariant 
partition function of an orbifold, Vafa \cite{v} realised that there is 
an ambiguity related to phase factors associated to the contributions of
the different twisted sectors of the partition function. It is possible
to weight the contributions of the $(g,h)$-twisted sector, where $g,h$ are
elements of the orbifold group $\Gamma$, by a phase $\beta_{g,h}$ without 
spoiling modular invariance. These phases are related to elements of 
$H^2(\Gamma,U(1))$ in such a way that if $\alpha_{g,h}$ is a 2-cocycle, 
the phases are of the form: $\beta_{g,h}=\alpha_{g,h}\alpha^{-1}_{h,g}$. 
These phases simply mean a freedom in changing the phase of some operators: 
the twisted fields in a sector twisted by $g$ pick up a phase $\beta_{g,h}$ 
when the element $h$ of the orbifold group is acting on them.

The effect of adding discrete torsion was analyzed in detail in some 
\ZNM\ heterotic orbifolds with standard embedding  \cite{fiq}. Some of 
these models have zero or negative Euler characteristic. These models 
are also related to some complete intersection CY manifolds.

Some examples were discused in detail by the authors of \cite{vw}. They 
found that the twisted fields in models with discrete torsion are
related to a deformation of the orbifold singularities (in contrast to
a blow-up of the singularities in the case without discrete torsion). 
However, some of the necessary deformations to completely smooth out the 
orbifold are absent, i.e.\ some singularities (conifolds) remain. 
As was commented by Vafa \cite{v}, the discrete torsion is related 
to some B-flux on a 2-cycle. The identification of this 2-cycle is carried 
out in \cite{am}. There, a relation was found between the discrete torsion 
and the torsion part of the homology of the target space 
(more precisely, the torsion part of the blow-up of the deformed orbifold).

Discrete torsion is implemented in theories containing $D$-branes by using 
projective representations of the orbifold group \cite{d,df}. These projective 
representations are classified by  $H^2(\Gamma,U(1))$ in complete analogy 
with the discrete torsion in closed string theories found in \cite{v}. 
Furthermore, the phases in projective representations appear in the amplitudes 
\cite{d} as a factor $\alpha_{g,h}\alpha^{-1}_{h,g}$ where $\alpha$ is the 
cocycle. Recently \cite{g} the relation between the closed string sector 
discrete torsion and the discrete torsion of the open string sector has been 
analyzed in disk amplitudes with a twisted closed state and a photon. In order 
for this amplitude to be invariant under $\Gamma$ the discrete torsion $\beta$ 
and the 2-cocycle must be related as above. $D$-brane charges are in 
correspondence with irreducible projective representations of $\Gamma$, i.e.\ 
for each irreducible representation there is a generator of the charge 
lattice. 

The deformation that was previously analyzed in the context of closed 
string theories \cite{vw} can now be seen as deformations of the 
superpotential, changing the F-flatness conditions. As in the closed string 
case, some singularities remain after switching on all the possible 
deformations (the remaining singularities are again conifolds). The relation 
of models of this type with the $AdS/CFT$ correspondence has been analyzed 
in \cite{beren}. 

The B-flux responsible for the discrete torsion parameter is related to some 
flux over a torsion 2-cycle. A similar problem has been analyzed in Type I 
strings: B-flux on the tori of an orbifold $T^4/\IZ_N$ \cite{kst,bps,ab}. 
There, the new features appear at the level of untwisted tadpoles (reduction 
of the rank) and at the level of the closed twisted sector (some extra tensor 
multiplets in six dimensions). These models are T-dual to models with 
non-commuting Wilson Lines \cite{k2,w,cpw}. We will find some of these 
features also in dealing with the discrete torsion in open string theories.

Another way of understanding the discrete torsion has been realized in 
\cite{s}. As an analogue of the lift of the action of the orbifold group 
$\Gamma$ to the gauge group (orbifold Wilson lines), one can understand the 
discrete torsion as an ambiguity in lifting the orbifold group action to 
another structure (gerbes). 1-gerbes are related to 2-forms, and  the action 
of the orbifold group on it is described in terms of the group cohomology 
$H^2(\Gamma,U(1))$.

The aim of this article is to understand discrete torsion in orientifold 
models.\footnote{A different approach has been taken by the authors of 
\cite{aaads}. They discuss a compact type I $\bZ_2\times\bZ_2$ orbifold. 
Their classification of possible models with unbroken supersymmetry 
agrees with ours.} We restrict ourselves to non-compact orientifold
constructions. The compact cases will be treated in a future publication
\cite{kr}. In section \ref{torsion}, we review some of the characteristic 
features of discrete torsion in closed string \cite{v} and open string 
theories \cite{d}. In section \ref{noncomporb}, we analyze the $D3$-brane 
systems at an orbifold singularity in the presence of discrete torsion 
\cite{d,df}. The closed string sector, the tadpole consistency conditions 
and some models are studied in detail. We analyze also the deformation of 
the $\IZ_2 \times \IZ_2$ theory with discrete torsion to get the usual 
$\IZ_2$ and conifold theories \cite{kw}. In section \ref{noncompori}, 
non-compact orientifolds are constructed with a set of $D3$-branes at 
the orbifold singularity. In general, some $D7$-branes are also needed in 
order to cancel the Klein bottle contribution to the tadpoles. Some 
consistency conditions are found in different sectors: the closed string 
sector, the open string sector and the tadpole cancellation conditions. In 
all the models we consider, these three conditions lead to the same 
restriction on the discrete torsion parameter $\eps$: only real values are 
allowed in the orientifold case, i.e. $\eps=\pm 1$. The orientifold involution 
is related to real projective representations. Four types of orientifold models
are found. This classification is based on the possibility of having vector
structure or no vector structure for each of the two generators of \ZNM. The 
open string spectrum and the tadpole conditions are different for each of the 
four cases. We determine explicitly the open string spectrum of the 
$D3$-branes. Some of the deformations of the superpotential, that in the 
orbifold case are allowed, are not present in the orientifold case. 
In particular, only the deformation to the $\IZ_2$ theory survives in the 
$\IZ_2 \times \IZ_2$ model.

Finally, some tools used in this paper are explained in the appendices. 
Many of the results of this article are based on the theory of of
complex and real projective representations \cite{karp,h}. In appendix
\ref{app_proj}, we therefore give a summary of their basic properties.
Appendix \ref{app_tadpole} contains a detailed computation of the tadpole 
conditions for \ZNM\ orientifolds. The relation between the shift formalism 
and the usual construction using matrices to determine the spectrum of
orientifold models is explained in appendix \ref{app_shift}. We show how
the shift formalism can be generalized to treat orientifolds with discrete
torsion.

\section{Discrete torsion}
\label{torsion}

In the original paper of Vafa \cite{v} discrete torsion appeared as
phase factors in the one-loop partition function of closed string 
orbifold theories.
If we denote by $Z(g,h)$ the contribution of the $(g,h)$-twisted 
sector\footnote{By this we mean the contribution from world-sheets that
are twisted by $g$ in their space direction and by $h$ in their time 
direction.}, then the total partition function $Z$ is given by
\be \label{partfunc}
Z={1\over|\Gamma|}\sum_{g,h\in\Gamma}\beta_{g,h}Z(g,h),
\ee
where $\Gamma$ is the orbifold group. The usual case without discrete
torsion corresponds to $\beta_{g,h}=1\ \forall\,g,h$. The discovery of
Vafa was that \nontriv phases are consistent with modular invariance
if they satisfy the following conditions:
\be \label{betacon}
\beta_{g,g}=1,\qquad \beta_{g,h}=\beta_{h,g}^{-1},\qquad
\beta_{g,hk}=\beta_{g,h}\beta_{g,k}\qquad \forall g,h,k\in\Gamma.
\ee
This implies (see appendix \ref{app_proj}) that these phases are of 
the form $\beta_{g,h}=\alpha_{g,h}\alpha_{h,g}^{-1}$, where $\alpha$
is a 2-cocycle of the group $\Gamma$. The possible discrete torsions are
therefore classified by the group cohomology $H^2(\Gamma,U(1))$.
We will only consider Abelian orbifold groups, i.e.\ $\Gamma$ is
(isomorphic to) a product of cyclic groups. Moreover, as the internal
space of the string models we want to discuss is complex three-dimensional,
all the possible cases can be reduced to $\Gamma=\IZ_N$ or
$\Gamma=\IZ_N\times\IZ_M$. It is known (see e.g.\ \cite{karp}) that 
$H^2(\IZ_N,U(1))=1$ and $H^2(\IZ_N\times\IZ_M,U(1))=\IZ_{\gcd(N,M)}$.
As a consequence, discrete torsion is only possible in
$\IZ_N\times\IZ_M$ orbifolds and is generated by one element of
$\IZ_{\gcd(N,M)}$. More precisely, let $g_1,g_2$ be the generators of
$\IZ_N,\IZ_M$ respectively, $p=\gcd(N,M)$ and choose 
$\beta_{g_1,g_2}=\omega_p^m$, with $\omega_p$ a $p$-th primtive 
root of unity and $m=1,\ldots,p$. All the other phases $\beta_{g,h}$ 
are then fixed by (\ref{betacon}). In the notation of appendix \ref{app_proj},
eq.\ (\ref{beta}), they read
\be \label{beta_gh}
\beta_{(a,b),(a^\prime,b^\prime)}=\eps^{ab^\prime-ba^\prime},
\qquad {\rm with\ }\eps=e^{2\pi im/p},\ m=1,\ldots,p.
\ee

As the partition function encodes the spectrum of the considered orbifold
model, it is clear that discrete torsion modifies the spectrum. From a 
geometrical point of view this change in the spectrum can be understood 
from the fact that discrete torsion changes the cohomology of the internal 
space \cite{vw}. However, the spectrum of the untwisted sector remains 
unchanged, as can be seen from $\beta_{g,e}=1$ (this follows from the third 
equation in (\ref{betacon}) by setting $h=g$ and $k=e$, where $e$ is the 
neutral element of $\Gamma$).

The generalization of discrete torsion to open strings has been found 
by Douglas \cite{d}. The matrices $\gamma_g$ that represent the action 
of the elements $g$ of the orbifold group $\Gamma$ on the Chan-Paton
indices of the open strings form a projective representation:
$\gamma_g\gamma_h=\alpha_{g,h}\gamma_{gh}$. The $\alpha_{g,h}$ are
arbitrary \nonze complex numbers. They are called the factor system
of the projective representation $\gamma$ and they form a 2-cocycle
in the sense that they satisfy (\ref{cocycle}).
Two matrices $\gamma_g$ and $\hat\gamma_g$ are considered
projectively equivalent if there exists a \nonze complex number $\rho_g$
such that $\hat\gamma_g=\rho_g\gamma_g$. As shown in appendix \ref{app_proj}
the set of equivalence classes of cocycles $\alpha$ is $H^2(\Gamma,U(1))$.
Thus, the ambiguity due to the projective representations in the open 
string sector is classified by the same group cohomology as the discrete
torsion of \cite{v} in the closed string sector. It is therefore natural
to assume that choosing a \nontriv factor system $\alpha_{g,h}$ corresponds
to discrete torsion in the open string sector. Moreover it was shown in
\cite{d,g} that a factor system $\alpha_{g,h}$ in the open string sector
of some orbifold model leads to phases 
$\beta_{g,h}=\alpha_{g,h}\alpha_{h,g}^{-1}$ in the closed string partition 
function of this model.

\section{Non-compact orbifold construction}
\label{noncomporb}

Let us consider a set of $D3$-branes at an orbifold singularity of the 
non-compact space $\IC^3/\Gamma$, where $\Gamma=\IZ_N\times\IZ_M$.
The effect of discrete torsion in such models has been studied in 
\cite{d,df,beren}. For completeness and to fix our notations we 
first summarize their results and then expand on them.

\subsection{Closed string spectrum}
\label{noncomcoh}
The closed string spectrum in \D4 can be obtained from the cohomology of the
internal space $\IC^3/\Gamma$. Strictly speaking, this spectrum is continuous 
because $\IC^3/\Gamma$ is non-compact. However, as noted in footnote 2 of 
\cite{dm}, it is interesting to determine the massless spectrum that would 
emerge, if the internal space were compactified. This is what we want to do.
The analysis is similar to the one performed by Vafa and Witten \cite{vw}.
The cohomology can be split in untwisted and twisted contributions:
$H^{p,q}=H^{p,q}_{\rm untw}+\sum_{g\in\Gamma\setminus\{e\}} H^{p,q}_g$.
The untwisted Hodge number $h^{p,q}_{\rm untw}=\dim(H^{p,q}_{\rm untw})$
is just given by the number of $\Gamma$-invariant $(p,q)$-forms on $\IC^3$.
This result is independent of the discrete torsion.
The twisted contributions are due to the singularities of $\IC^3/\Gamma$.
In the sector twisted by $g$, one has to find the $\Gamma$-invariant forms
that can be defined on the subspace $\cM_g$ of $\IC^3$ that is fixed under
the action of $g$. If discrete torsion is present, then the forms on 
$\cM_g$ must be invariant under the action of $h$ combined with a
multiplication by $\beta_{g,h}$ $\forall\,h\in\Gamma$. In the non-compact case 
that we are treating here, $\cM_g$ is either a point at the origin of $\IC^3$ 
or a complex plane located at the origin of the transverse $\IC^2$.
In the former case there is only a $(0,0)$-form. If no discrete torsion
is present, it is invariant and contributes one unit to $h_g^{1,1}$
or to $h_g^{2,2}$ depending on the specific model. In the case
with discrete torsion, there is no contribution to the cohomology
from this sector. If $\cM_g$ is a complex plane, then, in the case without
discrete torsion, the $(0,0)$-form and the $(1,1)$-form are invariant.
They contribute one unit to $h_g^{1,1}$ and to $h_g^{2,2}$. In the case
with discrete torsion the $(0,0)$-form and the $(1,1)$-form are invariant
if $\beta_{g,h}=1\ \forall\,h$. The $(0,1)$-form and the $(1,0)$-form
are invariant if the action of $h$ multiplies these forms by a phase which
is opposite to $\beta_{g,h}\ \forall\,h$. In this case, they contribute one 
unit to $h_g^{1,2}$ and to $h_g^{2,1}$.

For simplicity let us first take $\Gamma=\IZ_2\times\IZ_2$ \cite {d} and then
indicate the generalization to \ZNM. Using the method explained
above, the Hodge diamond of the untwisted cohomology is found to be

$$\matrix {& & & &1& & &  \cr
           & & &0& &0& &  \cr
           & &0& &3& &0&  \cr
           &1& &3& &3& &1 \cr
           & &0& &3& &0&  \cr
           & & &0& &0& &  \cr
           & & & &1& & &   }$$

For all the other \ZNM\ models, one finds nearly the same untwisted cohomology,
the only difference being that in those cases 
$h^{2,1}_{\rm untw}=h^{1,2}_{\rm untw}=0$.

There are three twisted sectors in the $\IZ_2\times\IZ_2$ corresponding to the
three \nontriv elements (0,1), (1,0), (1,1), where we used the notation
introduced in appendix \ref{app_proj} (below eq. (\ref{Htwo_rels})). 
Each of these sectors gives the same contribution to the twisted cohomology.  
In the case without discrete torsion:

$$\matrix {& & & &0& & &  \cr
           & & &0& &0& &  \cr
           & &0& &1& &0&  \cr
           &0& &0& &0& &0 \cr
           & &0& &1& &0&  \cr
           & & &0& &0& &  \cr
           & & & &0& & &   }$$

In the case with discrete torsion:

$$\matrix {& & & &0& & &  \cr
           & & &0& &0& &  \cr
           & &0& &0& &0&  \cr
           &0& &1& &1& &0 \cr
           & &0& &0& &0&  \cr
           & & &0& &0& &  \cr
           & & & &0& & &   }$$

\noindent
For a general \ZNM\ orbifold one finds that, only if minimal discrete torsion 
$\eps=\omega_{\gcd{(N,M)}}$ (i.e.\ $m=1$ in (\ref{beta_gh})) is present,
is there a \nonze contribution to 
$h_{\rm tw}^{1,2}=\sum_{g\in\Gamma\setminus\{e\}}h_g^{1,2}$.
Three cases can be distinguished:

i) $N=M\quad \Rightarrow\quad h_{\rm tw}^{1,2}=h_{\rm tw}^{2,1}=3$

ii) $N=\gcd(N,M)<M\quad \Rightarrow\quad h_{\rm tw}^{1,2}=h_{\rm tw}^{2,1}=2$

iii) $N\neq\gcd(N,M)\neq M\quad \Rightarrow\quad 
              h_{\rm tw}^{1,2}=h_{\rm tw}^{2,1}=0$

\vskip5mm 
In order to obtain the closed string spectrum in \D4, one has to dimensionally
reduce the massless spectrum of type IIB supergravity in \D{10}: the metric
$g^{(10)}$, the NSNS 2-form $B^{(10)}$, the dilaton $\phi^{(10)}$, the
RR-forms $C^{(10)}_0$, $C^{(10)}_2$, $C^{(10)}_4$. (We only give the bosons,
the fermions are related to them by supersymmetry.)
This is done by contracting their Lorentz indices with the differential 
forms of the internal space.
The resulting spectrum has \N2 supersymmetry in \D4. 
For a general configuration we get:
\begin{itemize}
\item $g^{(4)}$, $(h^{1,1}+2h^{2,1})$ scalars (from $g^{(10)}$)

\item a 2-form, $h^{1,1}$ scalars (from $B^{(10)}$)

\item a scalar (from $\phi^{(10)}$)

\item a scalar (from the $C^{(10)}_0$)

\item a 2-form, $h^{1,1}$ scalars (from $C^{(10)}_2$)

\item $(h^{2,1}+1)$ vectors, $h^{1,1}$ 2-forms (from $C^{(10)}_4$).
\end{itemize}

\noindent These fit into the following \N2 SUSY representations
(see e.g.\ \cite{lf}): 
\begin{itemize}
\item a gravity multiplet (consisting of $g^{(4)}$, a vector and fermions)
\item a double tensor multiplet (consisting of two 2-forms, two scalars and
      fermions) 
\item $h^{1,1}$ tensor multiplets (consisting of a 2-form, three scalars and
      fermions)
\item $h^{2,1}$ vector multiplets (consisting of a vector, two scalars and
      fermions)
\end{itemize}
 
\noindent For the $\IZ_2\times\IZ_2$ example we find (in \N2 multiplets):

- without discrete torsion: gravity, a double tensor, 3 vectors and
  6 tensors
                                                         
- with discrete torsion: gravity, a double tensor, 6 vectors and 
  3 tensors

\subsection{Tadpoles}

In the orbifold case, only oriented surfaces appear in computing the tadpoles. 
There is only one involving branes: the cylinder. As the space is non-compact 
all the tadpoles with an inverse dependence on the volume of the internal 
coordinates vanish. These are the untwisted ones and those corresponding
to twisted sectors that leave one complex plane fixed. So there is no 
restriction on the total number of branes and on $\Tr\gamma_g$ if $g$ has
a fixed plane. This means that, for example, in the $\IZ_2\times\IZ_2$ case 
there is no restriction from tadpoles (that is the case of ref. \cite{d}). 

In general, one can add $D7_i$-branes to the system of $D3$-branes at 
the singularity, where the index $i=1,2,3$ refers to the complex plane with 
Dirichlet conditions. Then the cylinder amplitude can be split into four 
sectors: $33$, $7_i7_i$, $7_i7_j$ and $7_i3$. It is convenient to denote
a group element $g=g_1^ng_2^m$ of $\Gamma=\IZ_N\times\IZ_M$ by the 2-vector 
$\bar k=(n,m)$. Similarly, we form 2-vectors of the components of the twist 
vectors $v_i$ and $w_i$ that represent the action of $g_1$ and $g_2$ on the 
$i$-th complex plane: $\bar v_i=(v_i,w_i)$. If we define 
$s_i=\sin(\pi\bar k\cdot\bar v_i)$, then the cylinder contribution 
is of the form (see appendix \ref{app_tadpole}):
\beq  \label{tad_cyl}
\cC = \sum_{\bar k=(1,1)}^{(N,M)} \frac{1}{8s_1s_2s_3} 
        \left[8s_1s_2s_3 \Tr\gamma_{\bar k,3} 
     +{\textstyle \sum^3_{i=1}}2s_i\Tr\gamma_{\bar k,7_i}\right]^2.
\eeq
As $D7$-branes are not needed for the consistency of the orbifold models, we
restrict ourselves to configurations without $D7$-branes. In this case the
tadpole conditions are summarized by:
\beq \label{tad_orb}
\Tr\gamma_g\ =\ 0,\qquad \hbox{if $g$ has no fixed planes.}
\eeq

As we have seen, discrete torsion appears in two ways: as phase factors in
the closed string partition function and as a \nontriv factor system in the
projective representation of the orbifold group on the Chan-Paton indices
of open strings. The former could only modify the Klein bottle diagram the 
latter could only affect the M\"obius strip. Both diagrams are not present 
in the orbifold case. Consequently, in this case, the tadpole conditions are 
not changed by discrete torsion.

Using formula (\ref{com_char}) for the characters of a projective
representation, it is easy to find solutions to the tadpole conditions
(\ref{tad_orb}). According to the results of appendix \ref{app_proj}, the 
matrix $\gamma_g$ (with $g=g_1^ag_2^b$) of a general projective 
representation with discrete torsion $\eps$ is of the form
\be \label{gamma}
\bigoplus_{k,l}(\sqrt\eps)^{-ab}\,(\omega_N^k\gamma_{g_1})^a\,
                           (\omega_M^l\gamma_{g_2})^b\,\otimes\,\one_{n_{kl}},
\ee
where $\gamma_{g_1}$ and $\gamma_{g_2}$ are given in (\ref{irrep}),
$k=0,\ldots,N/s-1$, $l=0,\ldots,M/s-1$. We recall that $\gamma_{g_{1/2}}$ 
are $(s\times s)$-matrices, where $s$ is the smallest positive integer, 
such that $\eps^s=1$. One can readily verify that the regular representation, 
i.e.\ $n_{kl}=s\ \forall\,k,l$, is a solution of (\ref{tad_orb}). But there 
are many more solutions. For example each set of $n_{kl}$ that either only 
depends on $k$ or only depends on $l$ (i.e.\ $n_{kl}=\tilde n_k\ \forall\,l$ 
or $n_{kl}=\tilde n_l\ \forall\,k$) is possible. 

To make some further statements, we need to specify how the generators $g_1$,
$g_2$ of \ZNM\ act on the three complex coordinates of the internal space. We
choose the following action:
\bea \label{def_twistvec}
g_1: &&(z_1,z_2,z_3)\ \to\ 
       (e^{2\pi iv_1}z_1,e^{2\pi iv_2}z_2,e^{2\pi iv_3}z_3) \nonumber\\
g_2: &&(z_1,z_2,z_3)\ \to\ 
       (e^{2\pi iw_1}z_1,e^{2\pi iw_2}z_2,e^{2\pi iw_3}z_3) \nonumber\\
{\rm with} &&v={1\over N}(1,-1,0),\qquad w={1\over M}(0,1,-1).
\eea
For $N=M$, all possible actions of $g_1$, $g_2$ can be brought to the
form (\ref{def_twistvec}) by permuting the elements of $\IZ_N\times\IZ_N$.
If $N\neq M$, one can choose different (non-equivalent) actions for 
$g_1,g_2$. We restrict ourselves to one alternative:
\bea \label{def_twistvec'}
g_1: &&(z_1,z_2,z_3)\ \to\ 
       (e^{2\pi iv_1}z_1,e^{2\pi iv_2}z_2,e^{2\pi iv_3}z_3) \nonumber\\
g_2: &&(z_1,z_2,z_3)\ \to\ (e^{2\pi iw^\prime_1}z_1,
            e^{2\pi iw^\prime_2}z_2,e^{2\pi iw^\prime_3}z_3) \nonumber\\
{\rm with} &&v={1\over N}(1,-1,0),\qquad w^\prime={1\over M}(1,-2,1).
\eea
To distinguish the two different actions, we will denote the orbifold
corresponding to the latter by $\IZ_N\times\IZ_M^\prime$.

Two interesting cases are (i) $N=M$ and (ii) $M$ a multiple of $N$.
In the first case, there is no tadpole condition for $s=N$ and $s=N/2$
(the only \nonvan characters correspond to group elements that have fixed
planes, see eq.\ (\ref{com_char})). So the first condition arises for $s=N/3$, 
which only for $N\ge6$
leads to an orbifold with discrete torsion. The condition that the $n_{kl}$
of (\ref{gamma}) have to satisfy for $N=M=6$ and $s=2$ is 
$\sum_{k,l=0}^2n_{kl}e^{2\pi i(k+2l)/3}=0$. As shown above, it is easy to
find solutions for the $n_{kl}$. For higher $N$ more conditions of this
type have to be satisfied. If $M$ is a multiple of $N$, then there are
\nonvan characters even in the case of minimal discrete torsion, $s=N$:
$\Tr\gamma_{(0,cN)}=N\sum_{l=0}^{M/N-1}n_{0,l}\,\omega_M^{lcN}$, where 
$c=1,\ldots,M/N-1$. However, the group element $(0,cN)$ has a fixed plane
if the group action is as in (\ref{def_twistvec}). Thus, only for the
$\IZ_N\times\IZ_M^\prime$ orbifold do we get a tadpole condition in the
case $s=N$. A solution to this condition is to take $n_{0,l}=n\ \forall\,l$.
If $M/N$ is a prime number, then this solution is unique, else different
choices for $n_{0,l}$ are possible.

\subsection{Open string spectrum}

Discrete torsion is implemented in the relation between the elements 
$\gamma_g$ of the representation of the group acting on the Chan-Paton 
matrices \cite{d}:
\be  \label{gamma_gh}
\gamma_{g} \gamma_{h}\ =\ \alpha_{g,h} \gamma_{gh}.
\ee

As in the case without discrete torsion, one gets the gauge fields 
$\lambda^{(0)}$ and matter fields $\lambda^{(i)}$, $i=1,2,3$, taking 
the solutions to the projections:
\be  \label{spec_cond}
\gamma^{-1}_{g} \lambda \gamma_{g}\ =\ r(g) \lambda,     
\ee
where $r(g)$ is the matrix that represents the action of $g$ on 
$\lambda=(\lambda^{(0)},\lambda^{(1)},\lambda^{(2)},\lambda^{(3)})$.

Let us first treat the $\IZ_2\times\IZ_2$ model of \cite{d} in detail and
then show how this is generalized to more complicated models.
The only \nontriv phases in (\ref{gamma_gh}) are
\be  \label{dt22}
\alpha_{g,h}\ =\ -\alpha_{h,g}\ =\ i,
\ee
where we have denoted the generators of the first and second $\IZ_2$ by
$g$ and $h$. This corresponds to a projective representation with discrete 
torsion $\eps=\alpha_{g,h}\alpha_{h,g}^{-1}=-1$. From (\ref{gamma}), we
find that the matrices are of the form:
\bea  \label{gam22}
\gamma_e &= &\one_2 \otimes \one_n, \\
\gamma_g &= &\sigma_3 \otimes \one_n, \nonumber\\
\gamma_h &= &\sigma_1 \otimes \one_n, \nonumber\\
\gamma_{gh} &=& \sigma_2 \otimes \one_n, \nonumber 
\eea
where $\sigma_i$ are the Pauli matrices and $n$ is an arbitrary parameter
(it counts the number of dynamical $D$-branes).

The solution to (\ref{spec_cond}) is given by
\be \label{sol22}
\lambda^{(0)}\ =\ \one_2 \otimes X,\qquad 
\lambda^{(i)}\ =\ \sigma_i \otimes Z_i,
\ee
where $X,Z_i$ are arbitrary $(n\times n)$-matrices. This corresponds
to gauge group $U(n)$ and three adjoint matter fields $Z_1$, $Z_2$, $Z_3$ 
in \N1 multiplets. 
The superpotential can be obtained in the usual way and reads:
\be  \label{sp22}
W\ =\ \Tr(Z_{1}Z_{2}Z_{3}+Z_{2}Z_{1}Z_{3}).
\ee

It has to be completed the by the deformations
\be  \label{deform}
\Delta W\ =\ \sum_{i=1}^{3} \zeta_i \Tr Z_i,
\ee
where $\zeta_i$ are the twisted modes from the closed string spectrum.

For general $\gamma$-matrices (\ref{gamma}), representing the action of \ZNM\ 
on the open strings, the solution to (\ref{spec_cond}) is a gauge group
\be  \label{ZNM_gauge}
\prod_{k=0}^{N/s-1}\prod_{l=0}^{M/s-1}U(n_{kl})
\ee
with matter in adjoint and bifundamental representations. For some models
the tadpole conditions impose restrictions on the numbers $n_{kl}$.
The spectrum can easiest be obtained from the corresponding quiver diagram,
as explained in \cite{beren}. A more rigorous way to find the spectrum uses
the shift formalism developed in \cite{afiv}. We explain this method in
appendix \ref{app_shift}. Note that a \ZNM\ orbifold with discrete torsion
characterized by the integer\footnote{Recall that $s$ was defined as the 
smallest positive integer, such that $\eps^s=1$.} $s$ has the same open 
string spectrum as the orbifold $\IZ_{N/s}\times\IZ_{M/s}$ without discrete
torsion. The only difference appears in the superpotential where some of
the terms acquire additional phases \cite{beren}.

In table \ref{noncomorb} we present the explicit solution for some models.
\begin{table}[pht!]
\renewcommand{\arraystretch}{1.25}
$$\begin{array}{|c|c|c|c|} \hline
  \Gamma &\rm torsion &\rm gauge\ group &\rm matter \\
\hline\hline
  \IZ_2 \times \IZ_2 &-1 &U(n) &3\ adj\\
\hline
  \IZ_3 \times \IZ_3 &e^{2\pi i/3} &U(n) &3\ adj\\
\hline
  \IZ_2 \times \IZ_4 &-1  &U(n_1) \times U(n_2) &(adj,\one) + (\one,adj)\\
   &&&+ 2(\fund,\antifund) +2(\antifund, \fund)  \\
\hline
  \IZ_4 \times \IZ_4 &i  &U(n) &3\ adj\\
\hline
  \IZ_4 \times \IZ_4 &-1  &U(n_1) \times U(n_2)  
   &(\fund,\antifund,\one,\one)+(\one,\fund,\antifund,\one)+
    (\one,\one,\fund,\antifund)\\
   &&\quad \times U(n_3) \times U(n_4) 
   &+(\fund,\one,\antifund,\one)+(\one,\fund,\one,\antifund)
    +(\fund,\one,\one,\antifund) \\
   &&&+{\rm\ conjugate}\\
\hline
  \IZ_2 \times \IZ_6 &-1 &U(n_1) \times U(n_2) \times U(n_3)
   &(\fund,\antifund,\one)+(\fund,\one,\antifund)+(\one,\fund,\antifund)\\
   &&&+(\antifund,\fund,\one)+(\antifund,\one,\fund)+(\one,\antifund,\fund)\\
   &&&+ (adj,\one,\one) + (\one,adj,\one) + (\one,\one,adj)\\
\hline
  \IZ_2 \times \IZ_6' &-1 &U(n) \times U(n) \times U(n) 
   &3(\antifund,\fund,\one)+3(\fund,\one,\antifund)+3(\one,\antifund,\fund)\\
\hline
  \IZ_3 \times \IZ_6 &e^{2\pi i/3}  &U(n_1) \times U(n_2) 
   &(adj,\one) + (\one,adj)\\
   &&&+ 2(\fund,\antifund) +2(\antifund,\fund)\\
\hline
  \IZ_6 \times \IZ_6 &e^{2\pi i/6}  &U(n) &3\ adj \\
\hline
  \IZ_6 \times \IZ_6 &e^{2\pi i/3}  &U(n_1) \times U(n_2) 
   &(\fund,\antifund,\one,\one)+(\one,\fund,\antifund,\one)+
    (\one,\one,\fund,\antifund)\\
   &&\quad \times U(n_3) \times U(n_4) 
   &+(\fund,\one,\antifund,\one)+(\one,\fund,\one,\antifund)
    +(\fund,\one,\one,\antifund) \\
   &&&+{\rm\ conjugate}\\
\hline
  \IZ_6 \times \IZ_6 &-1  &U(n_1)^3\times U(n_2)^3\times U(n_3)^3  
   &(\underline{\underline{\fund,\one,\one};
        \one,\one,\one;\underline{\antifund,\one,\one}})\\
   &&&+(\underline{\underline{\fund,\antifund,\one};
        \one,\one,\one;\one,\one,\one})\\
   &&&+(\underline{\underline{\fund,\one,\one};
     \underline{\one,\one,\antifund};\one,\one,\one})\\[1ex]
\hline
\end{array}$$
\caption{Open string spectrum of some non-compact orbifolds with discrete 
torsion. The conjugate representations are obtained exchanging fundamentals 
and antifundamentals. Underlining stands for all cyclic permutations of the
underlined elements (first line) respectively cyclic permutations of groups
of three entries (second line). If two groups of entries are underlined,
they are permuted simultaneously and not independently. Thus, in the case of 
$\bZ_6\times\bZ_6$, $\eps=-1$, one has 27 matter representations. 
\label{noncomorb}}
\end{table}

Note that the $\IZ_2\times \IZ_6'$ example with discrete torsion $\eps=-1$ 
would have \nonab gauge anomalies if the gauge group were 
$U(n_1)\times U(n_2)\times U(n_3)$, with $n_1\neq n_2\neq n_3$.
However, as mentioned in the previous subsection, the requirement of
tadpole cancellation implies that $n_1=n_2=n_3$.\footnote{The relation
between anomaly freedom and tadpole cancellation has been discussed in 
\cite{lr,abiu}.} This can be seen as follows.
The action of $\IZ_2\times \IZ_6'$ on the complex coordinates is as in 
(\ref{def_twistvec'}). There are three twisted sectors (we are not taking into 
account the inverse because they give the same tadpole cancellation condition)
that do not have any fixed plane: $h$, $h^2$ and $gh$, where $g$ and $h$ are
the generators of $\IZ_2$ and $\IZ_6^\prime$. 
Three tadpoles must be cancelled:
\be
\Tr\gamma_h = 0,\qquad \Tr\gamma_{gh} = 0,\qquad \Tr\gamma_{h^2} = 0. 
\ee

The action of the orbifold group on the Chan-Paton indices can be chosen as
\bea
\gamma_h &= &diag(\one_{n_1},-\one_{n_1},e^{2\pi i/6}\one_{n_2},
e^{8 \pi i/6}\one_{n_2},e^{4 \pi i/6}\one_{n_3},e^{10 \pi i/6}\one_{n_3}),
\nonumber\\
\gamma_g &= &
\left( \begin{array}{cccccc}
0 & 1& & & &  \\
1 & 0& & & &  \\
  &  &0&1& &  \\
  &  &1&0& & \\
  &  & & &0&1  \\
  &  & & &1&0  
\end{array} \right).
\eea

The first two tadpole conditions are inmediatly satisfied but the third one 
implies:
\be
n_1 + e^{2 \pi i/3} n_2 +e^{-2 \pi i/3} n_3 = 0.
\ee
The solution coincides with the condition for \nonab gauge anomaly 
cancellation: $n_1=n_2=n_3$.

A similar calculation shows that, in the case of $\IZ_6\times\IZ_6$
with discrete torsion $\eps = -1$, the gauge group is not of the most
general form (\ref{ZNM_gauge}). One needs $n_{1,1}=n_{1,2}=n_{1,3}$,
$n_{2,1}=n_{2,2}=n_{2,3}$ and $n_{3,1}=n_{3,2}=n_{3,3}$ for tadpole
cancellation. The same condition can also be deduced from the requirement
of anomaly freedom.

\subsection{Resolution of the singularities}

$D$-branes are interesting probes of  geometry and topology. World volume 
theories of $D$-branes at a singularity are related to the structure of this 
singularity. Spacetime can even be understood as a derived concept, emerging 
from \nontriv moduli spaces of the theory on the $D$-brane \cite{dm,dgm}.

In our case, there is a correspondence between the moduli space of the theory 
and the transverse space to the $D3$-branes. This correspondence allows a map 
between different theories and different singularities (this has been 
extensively studied, see e.g.\ \cite{mp,kw}).

Cases without discrete torsion are related to deformations of the D-flatness 
conditions. Fayet-Iliopoulos terms are controlled by the twisted sector moduli.
Non-trivial values of the Fayet-Iliopoulos parameters can be interpreted as 
resolutions of the singularities transverse to the branes. Arbitrary values of 
these terms resolve completely the singularity.

Cases with discrete torsion are related to deformations of the F-flatness 
conditions \cite {d,df}. The singularities are not smoothed out by blow-ups 
but by deformations \cite{vw}. However, these singularities cannot be 
completely smoothed out due to the lack of a sufficient number of twisted 
states. Some singularities are remaining.

Resolutions and deformations are the two main strategies of desingularization 
of orbifolds. An analysis of the different desingularizations associated to 
some orbifold singularities is done in \cite{j}. Only a small number of the 
possible desingularizations are available in string theory.

In this section we want to find the field theories related to the deformation 
of the $\IC /\IZ_2 \times \IZ_2$ singularity with discrete torsion $\eps=-1$. 
The analysis in the orientifold case will be derived from this one.
    
From the three adjoint fields one can construct the $SU(n)$ invariants. We 
consider the diagonal $U(1)$ as decoupled \cite{mr}. These $SU(n)$ 
invariants are of the form:
\beqa
M_{i,j} &= &\Tr(Z_{i}Z_{j}),\\
B       &= &\Tr(Z_{1}[Z_{2},Z_{3}]). \nonumber
\label{inv}
\eeqa
They satisfy the following relation:
\beq
B^2\ =\ \det M.
\label{inv2}
\ee

Using the F-flatness conditions coming from the superpotential (\ref{sp22}),
(\ref{deform}) above, one can obtain the following relation between the 
invariants:
\beq
\det 
\left( \begin{array}{ccc}
M_{11} & \zeta_3 & \zeta_2 \\
\zeta_3 & M_{22} & \zeta_1 \\
\zeta_2 & \zeta_1 & M_{33}
\end{array} \right)
\ =\ B^2.
\label{deter}
\eeq

There are three different singularities derived from this case:
\begin{itemize}
\item[(i)] If all the $\zeta_i$ are equal to zero the above equation between 
the invariants is of the form:
\beq
M_{11}M_{22}M_{33}\ =\ B^2.
\label{z2z2}
\eeq
This is a $\IZ_2 \times \IZ_2$ singularity.

\item[(ii)] If only one of the three $\zeta_i$, say $\zeta_1$, is different 
from zero, the equation relating the invariants is:
\beq
M_{11}(M_{22}M_{33}-\zeta_1 ^2)\ =\ B^2.
\label{z2}
\eeq
The singularity is at the points 
$(M_{22}M_{33}-\zeta_1 ^2,M_{11}M_{33},M_{11}M_{22},-2B)=(0,0,0,0)$. 
This corresponds to a $\IC^*$ of singularities. Changing the variables,
such that the singularity will pass through the origin, one has:
\beqa
M_{22} &= &\zeta_1 +x+y, \\
M_{33} &= &\zeta_1 +x-y. \nonumber
\label{z2cv}  
\eeqa
Taking the lowest order in the polynomial (just for analyzing the points near 
the singularity), gives:
\beq
M_{11}(2x \zeta_1 ^2)\ =\ B^2.
\label{z2cv2}  
\eeq
That is a $\IZ_2$ singularity. In order to obtain the field theory, 
let us give the following vev's to the fields:
\beq
\vev{Z_1}\ =\ 0,\qquad \vev{Z_2} \ =\ v \sigma_3,\qquad
\vev{Z_3} \ =\ v \sigma_3.
\label{vevz2}
\eeq
Plugging these values back into the F-flatness conditions, allows us to get 
the deformations:
\beq
\zeta_2\ =\ \zeta_3 = 0,\qquad \zeta_1\ =\ 2v^2.  
\label{defz2}
\eeq
These vev's break the group down to $U(n/2) \times U(n/2)$. The remaining 
spectrum is:
\begin{center}
\begin{tabular}{c|c|c}
 & $U(n/2)$ & $U(n/2)$ \\
\hline
$\phi_1$ & adj & 1 \\
$\phi_2$ & 1 & adj \\
$A_{12}$ & $\fund$ & $\antifund$ \\
$B_{12}$ & $\fund$ & $\antifund$ \\
$A_{21}$ & $\antifund$ & $\fund$ \\
$B_{21}$ & $\antifund$ & $\fund$ \\

\end{tabular}
\end{center}
The $A_{12}$, $B_{12}$, $A_{21}$, $B_{21}$, $\phi_1$ and $\phi_2$ can be
seen as the components of the $Z_i$ matrices:
\beq
Z_1 \ = \ 
\left( \begin{array}{cc}
B_{11} & B_{12} \\
B_{21} & B_{22} 
\end{array} \right), 
\label{zz21}
\eeq
\beq
Z_2 \ =\ v \sigma_3 +
\left( \begin{array}{cc}
(M_1 + \phi_1)/2 & 2A_{12} \\
2A_{21} & (M_2 + \phi_2)/2
\end{array} \right),
\label{zz22}
\eeq
\beq
Z_3 \ =\ v \sigma_3 +
\left( \begin{array}{cc}
(M_1 - \phi_1)/2 & -2A_{12} \\
-2A_{21} & (M_2 - \phi_2)/2
\end{array} \right).
\label{zz2}
\eeq
The fields $M_i$ and $B_{ii}$ get masses. The relation between the 
non-diagonal components of the $Z_2$ and $Z_3$ fields comes from the 
degrees of freedom that are eaten by the Higgs mechanism. The superpotential 
for the massless fields is obtained by integrating out the massive states:
\beq
W\ =\ -2 \Tr [(\phi_1+\phi_2) (A_{12}B_{21}+A_{21}B_{12})].
\label{wz2}
\eeq

\item[(iii)] Let us now take all the three $\zeta_i$ different from zero. 
For that, one can analyze how are the possible deformations of the $\IZ_2$ 
case above (another possibility with the same result is taking the vev's of 
the fields all proportional to $\sigma_3$). Let us take for simplicity 
$U(1) \times U(1)$ as the gauge group at the $\IZ_2$ singularity.
The deformations can be expressed in terms of the massless fields using the 
F-flatness conditions. There are four possible solutions:
\beq
\Delta W\ =\ \pm 2 v^2  [4v + \frac{8}{v} A_{12}A_{21} 
            +\frac{1}{4v}(\phi_1^2 +\phi_2^2)],
\label{dc1}
\eeq
\beq
\Delta W\ =\ \pm 2 v^2  \frac{1}{4v}(\phi_1^2 -\phi_2^2).
\label{dc2}
\eeq
The second one is the deformation considered by Klebanov and Witten \cite{kw} 
that takes to the theory at the conifold. The superpotential 
gives masses to the two adjoints, leaving as massless spectrum:
\begin{center}
\begin{tabular}{c|c|c}
 & $U(n/2)$ & $U(n/2)$ \\
\hline

$A_{12}$ & $\fund$ & $\antifund$ \\
$B_{12}$ & $\fund$ & $\antifund$ \\
$A_{21}$ & $\antifund$ & $\fund$ \\
$B_{21}$ & $\antifund$ & $\fund$ \\

\end{tabular}
\end{center}

\end{itemize}

\section{Non-compact orientifold construction}
\label{noncompori}

Let us locate a set of $D3$-branes at an orbifold singularity of 
$\IC^3/\Gamma$. 
The action of the orbifold group $\Gamma$ permutes the branes in an 
appropriate way \cite{dm}. We take an orientifold of the form 
$\Omega'=\Omega (-1)^{F_L}R_1 R_2 R_3$ in order to preserve \N1 supersymmetry 
in the effective theory in four dimensions (A models in \cite{pru}). 
The action of $\Omega$ on the Chan-Paton matrices can be chosen to be either 
symmetric (A1 models) or antisymmetric (A2 models). In general these models 
will need the presence of $D7$-branes for consistency. This can be deduced 
from the tadpole cancellation conditions.

\subsection{Closed string spectrum}

As in the orbifold case, the closed string spectrum is continuous because
the space $\IC^3/\Gamma$ is non-compact. Again, the philosophy is to find
the spectrum that would exist if the internal space were compactified.
The $\Omega$ parity is accompanied by a $J$ operation that relates states
from one twisted sector with states from the inverse sector.
In order to obtain the  $\Omega J$ invariant states one must 
combine the states from the $g$ and $g^{-1}$ twisted sectors. Because of this, 
the counting of states is different compared to the orbifold case.

For sectors that are invariant under the $J$ permutation, i.e.\ the untwisted 
sector and the order-two twisted sectors, the spectrum in four dimensions is
obtained by dimensionally reducing the fields of the type I string in ten 
dimensions. These fields are the metric $g^{(10)}$ and the dilaton 
$\phi^{(10)}$ in the NSNS sector and the 2-form $C_2^{(10)}$ in the RR sector.
Their Lorentz indices have to be contracted with the differential forms of
the untwisted and order-two twisted cohomology (see section \ref{noncomcoh}). 
In \N1 representations these sectors give: the gravity multiplet, a linear 
multiplet and $(h^{1,1}+h^{2,1})_{\rm untw+\hbox{\scriptsize order-two}}$ 
chiral multiplets.

One can split the remaining sectors into two types: 

- The sectors twisted by $g$ that give the same contribution to the cohomology
  as their inverse sectors, i.e.\ twisted by $g^{-1}$ (these sectors always
  have fixed planes). 
  Together they give $h_g^{1,1}$ linear multiplets, $h_g^{2,1}$ vectors and 
  $h_g^{1,1}+h_g^{2,1}$ chiral multiplets \cite{klein}.

- The sectors twisted by $g$ that give a different contribution to the 
  cohomology as their inverse sectors. Only one combination 
  of the fields survives: $h_g^{1,1}$ linear multiplets and  $h_g^{2,1}$ 
  vectors. If the contribution from the $g$-twisted sector to $h_g^{1,2}$ 
  and  $h_g^{2,1}$ is not the same (that happens if the discrete torsion is 
  different from $\pm 1$), one cannot match properly the states from one 
  sector to the inverse sector. An orientifold model that contains such a
  sector is ill-defined. This consistency criterium agrees with the one 
  we will find in the open string sector. There, it turns out that only 
  real values of the discrete torsion are allowed.\footnote{We thank Angel
  Uranga for pointing out to us that this can also be seen by noticing 
  that $\Omega J$ is not a symmetry of type IIB theory
  in the case of non-real discrete torsion.}

In table \ref{illdef}, one can compare the  $\IZ_2 \times \IZ_2$ and 
$\IZ_3 \times \IZ_3$ closed string spectrum. The first one admits 
the values $\eps=\pm 1$ for the discrete torsion. Note that, although the
cohomology is different in both cases, the closed string spectrum is the same
for both values of the discrete torsion. The $\IZ_3 \times \IZ_3$ model has 
$H^2(\Gamma,U(1))= \IZ_3$, that is $\eps= e^{\pm2\pi i/3}$ or $\eps=1$. 
Only the last value is allowed in the orientifold case.

\begin{table}[pht!]
\renewcommand{\arraystretch}{1.25}
\begin{center}
\begin{tabular}{|c|c|c|}
\hline
$\Gamma$ & torsion & closed string spectrum \\
\hline\hline
$\IZ_2 \times \IZ_2$ & 1 & 1 $G$ + 1 $L$ + 9 $\Phi$\\
& -1 & 1 $G$ + 1 $L$ + 9 $\Phi$ \\
\hline
$\IZ_3 \times \IZ_3$ & 1 &  1 $G$ + 5 $L$ + 6 $\Phi$ \\
& $e^{\pm2\pi i/3}$ & not well defined  \\[1ex]
\hline
\end{tabular}
\end{center}
\caption{
Closed string spectrum for the orientifolds $\bZ_2 \times \bZ_2$ and
$\bZ_3 \times \bZ_3$ with and without discrete torsion . 
The spectrum is organized in \N1 representations ($G$ is the gravitational 
multiplet, $L$ a linear multiplet and  $\Phi$ a chiral multiplet). 
\label{illdef} }
\end{table}

Let us explain in detail why the closed string spectrum cannot be 
consistently defined if $h_g^{2,1}\neq h_g^{1,2}$ for a sector twisted by $g$.
The states corresponding to this sector can be obtained from the shift 
formalism.\footnote{This will be explained in detail in \cite{kr}.} 
Take one sector $g$ with $h_g^{1,2}\neq0$ 
and  $h_g^{2,1}=0$. This gives $h_g^{1,2}$ helicity components $+1$ of some 
vectors and none with helicity $-1$. In the inverse sector $g^{-1}$ one gets  
$h_{g^{-1}}^{1,2}=0$ and $h_{g^{-1}}^{2,1}=h_g^{1,2}$. 
This leads to $h_{g^{-1}}^{2,1}$ helicity components $-1$ of some vectors. 
In type IIB theory this causes no problem, we get $h_g^{1,2}$ vectors. 
The problem arises if we want to impose the $J$ projection that relates the  
$g$-twisted sector with its inverse: One cannot match the states from these 
two sectors because they have opposite helicity.  

This rules out several orientifold models: $\IZ_3 \times \IZ_3$ with
discrete torsion,  $\IZ_4 \times \IZ_4$ with discrete torsion 
$\pm i$, $\IZ_3 \times \IZ_6$ with discrete torsion, etc. 
One can check that the orientifold projection is compatible with
discrete torsion only for the values $\eps=+1$ and $\eps=-1$. This can be 
seen from the formulae for $h_g^{1,2}$ and $h_g^{2,1}$ in the compact 
cases \cite{kr}:
\bea
h_g^{1,2} &= &{1\over|\Gamma|} \sum_{h\in\Gamma} \beta_{g,h}  
               \widetilde\chi(g,h) e^{2 \pi iv_h(g)},  \\
h_g^{2,1} &= &{1\over|\Gamma|} \sum_{h\in\Gamma} \beta_{g,h}
               \widetilde\chi(g,h) e^{-2 \pi iv_h(g)}. \nonumber   
\eea
Only if the discrete torsion is real,  are $h_g^{1,2}$ and $h_g^{2,1}$ 
equal for any twist. We will see that this
criterium in the closed string sector coincides with the one coming from 
the open string sector related to projective representations.

\subsection{Tadpoles} 

As stated above, discrete torsion appears as a phase between different twisted 
sectors in closed string theories and as a phase between the matrices that 
represent the action of $\Gamma$ on the Chan-Paton indices in the open string
sector. In order to compute the tadpole contribution, one must sum over three 
different diagrams: the cylinder ($\cC$), the M\"obius strip ($\cM$) and the 
Klein bottle ($\cK$). The first two diagrams contain a trace over the 
projective representation $\gamma$. Discrete torsion phases appear in the
M\"obius strip and in the Klein bottle. The three diagrams can be written as 
a sum over twisted sectors:
\bea \label{CMK}
&&\cC=\sum_{g\in\Gamma}\tilde\cC(g),\qquad
  \cM=\sum_{g\in\Gamma}\tilde\cM(g),\qquad
  \cK=\sum_{g,h\in\Gamma}\tilde\eps_h\,\beta_{h,g}\,\tilde\cK_h(g),\\
&&{\rm with}\quad \tilde\cC(g)\propto(\Tr\gamma_g)^2\quad{\rm and}\quad 
  \tilde\cM(g)\propto\Tr(\gamma_{\Omega g}^{-1}\gamma_{\Omega g}^\top).
\nonumber
\eea
The phases $\beta_{h,g}$ satisfy (\ref{betacon}) and are related to the 
discrete torsion. The additional factor $\tilde\eps_h=\pm1$ appears 
because of the possibility of having vector structure or not in each of 
the two factors of \ZNM. As $J$ relates the orbifold elements $h$ and 
$h^{-1}$, the only \nonvan contributions to the Klein bottle are 
(see appendix of \cite{afiv}) $\tilde\cK_e(g)$ and $\tilde\cK_{h^{(2)}}(g)$, 
where $e$ is the neutral element of $\Gamma$ and $h^{(2)}$ is of order two, 
i.e.\ $(h^{(2)})^2=e$. The first of these two contributions is not modified 
by discrete torsion, because (\ref{betacon}) implies $\beta_{e,g}=1$. 
The second contribution depends on discrete torsion, as explained in 
appendix \ref{app_tadpole}.

In general, models of this type require the presence of $D7$-branes. The 
tadpole calculation is detailed in appendix \ref{app_tadpole}. To give
the result of this calculation, we first introduce some notation. It is 
convenient to denote the group elements of $\Gamma=\IZ_N\times\IZ_M$ by 
$\bar k=(n,m)$ as in eq.\ (\ref{tad_cyl}). One has to distinguish even 
twists $\bar k$, i.e.\ there is a $\bar k'\in\Gamma$, such that 
$\bar k=2\bar k'$, from odd twists that cannot be written in this form.
If $N$ is odd and $M$ is even, there exists one order-two element $\bar k_1$.
If $N$ and $M$ are both even there are three order-two elements $\bar k_1$,
$\bar k_2$, $\bar k_3$, where the index denotes the complex plane that is
fixed by this element (i.e.\ $\bar k_1=(0,M/2)$, etc). As will be explained
below, the generalization of the notion of vector structure to \ZNM\ 
orientifolds leads to four different boundary conditions on the $\gamma$
matrices: $\gamma_{g_1}^N=\pm\one$, $\gamma_{g_2}^M=\pm\one$. We define the
numbers $\mu_i=\pm1$ (the index $i$ refers to the complex plane that is
fixed by the corresponding group element) by
\be  \label{def_mu}
\gamma_{g_1}^N=\mu_3\,\one,\quad 
\gamma_{g_2}^M=\mu_1\,\one,\quad
(\gamma_{g_1}\gamma_{g_2})^{{\rm lcm}(N,M)}=\mu_2\one.
\ee
Of course, only two of the $\mu_i$ are independent. One can show that
\be  \label{mu2}
\mu_2\ =\ \left\{\begin{array}{ll}
           \eps^{-MN/4}\,\mu_3\,\mu_1\ &{\rm if\ }{M\over\gcd(N,M)}{\rm\ and\ }
                                        {N\over\gcd(N,M)}{\rm\ odd}\\
           \mu_1                 &{\rm if\ }{M\over\gcd(N,M)}{\rm\ even}\\
           \mu_3                 &{\rm if\ }{N\over\gcd(N,M)}{\rm\ even}
           \end{array}\right.,
\ee
where $\eps$ is the discrete torsion parameter of (\ref{beta_gh}) and we
used that $\eps$ only takes the values $\pm1$.
Due to the fact that the $\gamma$-matrices form a {\em projective} 
representation, there may appear a phase $\delta_{\bar k}$ in the
M\"obius strip:
\be  \label{delta_k}
\Tr(\gamma_{\Omega\bar k}^{-1}\gamma_{\Omega\bar k}^\top)
=\delta_{\bar k}\Tr(\gamma_{\bar k}^2).
\ee
Finally, we define 
\be  \label{sincos_def}
s_i=\sin(\pi\bar k\cdot\bar v_i),\quad
\tilde s_i=\sin(2\pi\bar k\cdot\bar v_i),\quad
c_i=\cos(\pi\bar k\cdot\bar v_i).
\ee
In this notation the tadpole conditions for a general \ZNM\ orientifold read:
\begin{itemize}
\item[a)] If $N$ and $M$ are both odd:
  \be  \label{tad_NModd}
    8 \tilde{s}_1\tilde{s}_2\tilde{s}_3\Tr{\gamma_{2\bar k,3}} 
    + \sum_{i=1}^3 2 \tilde{s}_i \Tr{\gamma_{2\bar k,{7_i}}}
    = \delta_{\bar k} 32 s_1s_2s_3.
  \ee

\item[b)] If $N$ is  odd and $M$ is even:
  \begin{itemize}
  \item[-] for odd $\bar k$:
     \be \label{tad_cyl_Nodd}
       8 s_1s_2s_3\Tr{\gamma_{\bar k,3}} 
       + \sum_{i=1}^3 2 s_i \Tr{\gamma_{\bar k,{7_i}}} = 0,
     \ee
   \item[-] for even twists $2\bar k$:
     \be  \label{tad_Nodd}
       8 \tilde{s}_1\tilde{s}_2\tilde{s}_3\Tr{\gamma_{2\bar k,3}} 
       + \sum_{i=1}^3 2 \tilde{s}_i \Tr{\gamma_{2\bar k,{7_i}}}
       = \delta_{\bar k} 32 (s_1s_2s_3 - \tilde\eps_1 s_1 c_2 c_3),
     \ee
  where $\tilde\eps_1=\tilde\eps_{\bar k_1}$.
  \end{itemize}

  As explained in appendix \ref{app_tadpole}, in order to factorize the 
  amplitudes in the appropriate way, one must impose some restrictions 
  on the `vector structures'. In the present case, one needs\footnote{Strictly
  speaking, these relations only follow from tadpole factorization if $M>2$. 
  However, these conditions are still valid in the case $M=2$. This can be
  seen by similar arguments as the ones used in \cite{bl}. The same applies
  to the relations (\ref{vec_cond2}) below.}:
  \beq  \label{vec_cond}
  {\mu_1}^3 = {\mu_1}^{7_1} = -{\mu_1}^{7_2} = -{\mu_1}^{7_3} 
  = \tilde\eps_1.
  \eeq

\item[c)] If $N$ and $M$ are both even:\\
  Only in this case discrete torsion $\eps=-1$ is possible. The tadpole 
  conditions are:
  \begin{itemize}
  \item[-] for odd $\bar k$:
     \be \label{tad_cyl_NMeven}
       8 s_1s_2s_3\Tr{\gamma_{\bar k,3}} 
       + \sum_{i=1}^3 2 s_i \Tr{\gamma_{\bar k,{7_i}}} = 0,
     \ee
  \item[-] for even twists $2\bar k$:
     \be  \label{tadNMeven}
       8 \tilde{s}_1\tilde{s}_2\tilde{s}_3\Tr{\gamma_{2\bar k,3}} 
       + \sum_{i=1}^3 2 \tilde{s}_i \Tr{\gamma_{2\bar k,{7_i}}}
       = \eps^{-k_1k_2} \delta_{\bar k} 32 (s_1s_2s_3 
            - \sum_{i\neq j\neq k} \tilde\epsilon_i \beta_i s_i c_j c_k),
     \ee
     where $\tilde\eps_i=\tilde\eps_{\bar k_i}$,
     $\beta_i=\beta_{\bar k_i,\bar k}$.
  \end{itemize}
  As in the previous case, there are some conditions on the 
  `vector structures':
  \beqa  \label{vec_cond2}
    {\mu_1}^3 = {\mu_1}^{7_1} = -{\mu_1}^{7_2} = -{\mu_1}^{7_3} 
    = \tilde\eps_1, \nonumber\\
    {\mu_3}^3 = -{\mu_3}^{7_1} = -{\mu_3}^{7_2} = {\mu_3}^{7_3}
    = \tilde\eps_3.  
   \eeqa
   In addition the $\tilde\eps_i$ are related by
   \be  \label{eps_cond}
      \tilde\eps_1\tilde\eps_2\tilde\eps_3=\eps^{-MN/4}.
   \ee
\end{itemize}

Let us discuss some examples:
\begin{itemize}
\item $\IZ_2\times\IZ_2$: All tadpole conditions vanish. As in the orbifold
case, the contribution from the untwisted sector vanishes because the 
space $\IC^3/\Gamma$ transverse to the $D3$-branes is non-compact and the 
tadpoles depend on the inverse volume. The twisted sectors $(1,0)$, $(0,1)$ 
and $(1,1)$ each leave one complex plane fixed (the third, the first and the 
second respectively). The sum over the windings along the fixed direction
leads to a dependence on the inverse volume of the fixed plane.
In particular, no $D7$-branes are needed for this model to be consistent.
 
\item $\IZ_3\times\IZ_3$: There is one twisted sector that does not fix any 
plane but has only a fixed point at the origin, namely the sector twisted
by $g=(1,2)$ or by its inverse $g^{-1}=(2,1)$. Tadpole cancellation implies 
\cite{z,ks}:
\beq
    -\Tr{\gamma_{(1,2),3}} 
    + \frac13 (\Tr{\gamma_{(1,2),{7_1}}} + \Tr{\gamma_{(1,2),{7_2}}}
               - \Tr{\gamma_{(1,2),{7_3}}})
    \ = \ 4\delta_{(2,1)}.
\eeq
But this condition cannot be satisfied if there is discrete torsion. The only
allowed values of the discrete torsion parameter different from one are 
$e^{\pm 2\pi i/3}$. For these values there is a unique 
projective representation (see appendix \ref{app_proj}) with the following 
character:
\beq \label{Trg0}
\Tr \gamma_{g} \ =\ 0\qquad\forall\,g\neq e.
\eeq
($\Tr\gamma_e$ is related to the total number of branes.) Thus, we see that
the above condition can never be satisfied for this projective representation.
This inconsistency agrees with the one we found in the closed string sector of
the same model. 

A similar analysis can be done for other orientifolds, with the result that no
solution can be found for certain values of the discrete torsion. In 
$\IZ_N\times\IZ_N$ models \cite{df}, with discrete torsion $\eps=e^{2\pi im/N}$
and $m$ and $N$ coprime, there is only one irreducible representation%
\footnote{In general there are $N^2/s^2$ different irreducible 
representations. If, however, $\gcd(N,m)=1$, then the minimal positive 
integer $s$, such that $\eps^s=1$,
is $s=N$.} of the discrete group with this torsion as a factor system. This
means that the caracters of this representation vanish for all group elements
except for the unit element, i.e.\ (\ref{Trg0}) is valid for all of these 
models. Following the same argument as in the $\IZ_3\times\IZ_3$ case, one 
finds that all $\IZ_N\times\IZ_N$, $N>2$, orientifolds with minimal discrete 
torsion (i.e.\ $s=N$) are inconsistent. Again, this result agrees with the 
restrictions found in the closed string sector.

\item$\IZ_2\times\IZ_4$: The tadpole conditions are:
$$\renewcommand{\arraystretch}{1.25} \ba{l|l|l}
\bar k &\bar k\cdot\bar v &\hbox{tadpole condition}\\ \hline
(0,1) &\frac14(0,1,-1) &\Tr\gamma_{(0,1),7_2}-\Tr\gamma_{(0,1),7_3}=0\\
(0,3) &\frac34(0,1,-1) &\Tr\gamma_{(0,3),7_2}-\Tr\gamma_{(0,3),7_3}=0\\
(1,0) &\frac12(1,-1,0) &\Tr\gamma_{(1,0),7_1}-\Tr\gamma_{(1,0),7_2}=0\\
(1,1) &\frac14(2,-1,-1) &4\Tr\gamma_{(1,1),3}+2\Tr\gamma_{(1,1),7_1}
             -\sqrt2\Tr\gamma_{(1,1),7_2}-\sqrt2\Tr\gamma_{(1,1),7_3}=0\\
(1,2) &\frac12(1,0,-1) &\Tr\gamma_{(1,2),7_1}-\Tr\gamma_{(1,2),7_3}=0\\
(1,3) &\frac14(2,1,-3) &-4\Tr\gamma_{(1,3),3}+2\Tr\gamma_{(1,3),7_1}
             +\sqrt2\Tr\gamma_{(1,3),7_2}-\sqrt2\Tr\gamma_{(1,3),7_3}=0\\
(0,2) &\frac12(0,1,-1) &2\Tr\gamma_{(0,2),7_2}-2\Tr\gamma_{(0,2),7_3}
                 = -16\delta_{(0,1)}(\tilde\eps_2\beta_2-\tilde\eps_3\beta_3)
\ea
$$
There are two cases:\\
(i) If $\tilde\eps_2\beta_2=\tilde\eps_3\beta_3$, then the Klein bottle 
contribution in the $(0,2)$ sector vanishes. This means that no $D7$-branes 
are needed to cancel the tadpoles. This agrees with anomaly cancellation in 
the 33 sector. Using $\beta_2=\beta_{(1,2),(0,1)}=\eps$, 
$\beta_3=\beta_{(1,0),(0,1)}=\eps$ and eqs.\ (\ref{vec_cond2}), 
(\ref{eps_cond}), we find that $\tilde\eps_2\beta_2=\tilde\eps_3\beta_3$ 
implies ${\mu_1}^3{\mu_3}^3={\mu_3}^3$. Therefore this case is characterized 
by ${\mu_1}^3=+1$.\\
(ii) If $\tilde\eps_2\beta_2=-\tilde\eps_3\beta_3$, then $D7$-branes 
are present. In the case with discrete torsion $\eps=-1$, a minimal 
choice to satisfy the tadpole conditions consists in a set of $D7_2$-branes. 
Then all the traces vanish, except for the $(0,2)$ sector, as can be seen 
from (\ref{real_char}). With the minimal choice, one has
\be  \label{tad24}
 \Tr\gamma_{(0,2),7_2}=16\,\delta_{(0,1)}\tilde\eps_3\beta_3
                      =-16\,\delta_{(0,1)}{\mu_3}^3,
\ee
where we used (\ref{vec_cond2}) and $\beta_3=\beta_{(1,0),(0,1)}=-1$.
The relation $\tilde\eps_2\beta_2=-\tilde\eps_3\beta_3$ implies
${\mu_1}^3{\mu_3}^3=-{\mu_3}^3$ and thus ${\mu_1}^3=-1$. 
From (\ref{vec_cond2}) we then find that this case is characterized 
by ${\mu_1}^3=-{\mu_1}^{7_2}=-1$.

\end{itemize}

\subsection{open string spectrum}
An element $g$ of the orbifold group $\Gamma$ and the world-sheet parity
$\Omega$ act on the Chan-Paton matrices $\lambda$ as
\beq \label{action_gOm}
g:\ \lambda\ \longrightarrow\ \gamma_g\lambda\gamma_g^{-1},\qquad\qquad
\Omega:\ \lambda\ \longrightarrow\ \gamma_\Omega\lambda^\top\gamma_\Omega^{-1}.
\eeq
The open string spectrum, i.e.\ gauge fields $\lambda^{(0)}$ and matter 
fields $\lambda^{(i)}$, is obtained by taking the solutions to the 
projections:
\be  \label{ospec_cond}
\gamma^{-1}_{g} \lambda \gamma_{g}\ =\ r(g) \lambda,\qquad
\gamma^{-1}_{\Omega} \lambda^\top \gamma_{\Omega}\ =\ r(\Omega) \lambda,     
\ee
where $r(g)$ (resp.\ $r(\Omega)$) is the matrix that represents the action 
of $g$ (resp.\ $\Omega$) on 
$\lambda=(\lambda^{(0)},\lambda^{(1)},\lambda^{(2)},\lambda^{(3)})$.

The relation $\Omega^2=e$ gives a restriction on the matrix $\gamma_\Omega$:
\beq \label{gamom_cond}
\gamma_\Omega\,(\gamma_\Omega^{-1})^\top=c\,\one,
\eeq
where the constant $c$ can only take the values $\pm1$, i.e.\ 
$\gamma_\Omega$ is either symmetric or antisymmetric. A further condition
follows from $\Omega g\Omega=g$, which, using (\ref{action_gOm}), translates
to:
\beq \label{gamomg_cond}
\gamma_\Omega(\gamma_g^{-1})^\top (\gamma_\Omega^{-1})^\top\ =\ 
  \delta_g\, \gamma_g, 
\eeq
where the phase $\delta_g$ appears because the representation is only 
projective. More precisely, in the notation of appendix \ref{app_proj},
$\delta_g=c\,\beta_{\Omega,g}$. This can be seen as follows. Because of 
the special action of $\Omega$ on $\lambda$, eq.\ (\ref{action_gOm}), 
the matrices representing $\Omega g$ and $g\Omega$ are given by
\be  \label{gam_omg}
\gamma_{\Omega g}=\alpha_{\Omega,g}^{-1}\,\gamma_\Omega(\gamma_g^{-1})^\top,
\qquad \gamma_{g\Omega}=\alpha_{g,\Omega}^{-1}\,\gamma_g\gamma_\Omega.
\ee
From $\Omega g=g\Omega$ and (\ref{gamom_cond}), we find 
$\delta_g=c\,\alpha_{\Omega,g}\alpha_{g,\Omega}^{-1}$.
It is easy to see that the phase $\delta_g$ coincides with the one
defined in (\ref{delta_k}).

As we are interested in unitary projective representations,
(\ref{gamomg_cond}) can be transformed into
\beq \label{gammareal}
\gamma_\Omega\gamma_g^*\gamma_\Omega^{-1} \ =\  c\,\delta_g\,\gamma_g, 
\eeq
where we used $\gamma_\Omega^\top=c\,\gamma_\Omega$.
The world-sheet parity $\Omega$ relates one representation with its complex 
conjugate. This means that the matrices $\gamma_g$ of the orientifold form
a {\em (pseudo-)real} projective representation of the orbifold group. As 
the complex projective representations are classified by $H^2(\Gamma, \IC^*) 
= H^2(\Gamma,U(1))$, the real projective representations are classified 
by  $H^2(\Gamma, \IR^*) = H^2(\Gamma,\IZ_2)$. As a consequence, the
discrete torsion paramter $\eps$ can only take the real values $\pm 1$.

Let us make one further remark concerning the phases  $\beta_{\Omega,g}$.
They do not satisfy (\ref{beta_cond}) but rather 
$\beta_{\Omega,gh}=(\alpha_{g,h})^2\beta_{\Omega,g}\beta_{\Omega,h}$.
Moreover, one can always find an equivalent set of $\gamma$-matrices with
$\beta_{\Omega,g}=1\ \forall\,g\in\Gamma$ by defining 
$\hat\gamma_g=\sqrt{\beta_{\Omega,g}}\,\gamma_g$. If we choose the factor 
system $\alpha_{g,h}$ of the orbifold group $\Gamma$ as in (\ref{orbi_fs}), 
then $(\alpha_{g,h})^2=1\ \forall\,g,h$ and therefore the $\hat\gamma_g$ 
have the same factor system as the $\gamma_g$: $\hat\alpha_{g,h}=
\sqrt{\beta_{\Omega,g}\beta_{\Omega,h}\beta_{\Omega,gh}^{-1}}\,\alpha_{g,h}
=\alpha_{g,h}$. Thus, to fix the factor system of the orientifold completely,
one has to add one relation to (\ref{orbi_fs}):
\bea  \label{orien_fs}
\alpha_{g_1^a,g_2^b}=\alpha_{g_1,g_1^a}=\alpha_{g_2,g_2^b} &= &1, \qquad
a=1,\ldots,N,\quad b=1,\ldots,M, \nonumber\\
\delta_g &= &c \ \qquad\forall\,g\in\Gamma.
\eea
In the following, we will mostly leave $\alpha_{g,h}$, $\delta_g$ arbitrary. 
But to perform calculations, we choose the factor system (\ref{orien_fs}).

The reality condition (\ref{gammareal}) has an interesting consequence
concerning the classification of $\IZ_N$ orientifolds. Note, that for a
complex projective representation of $\IZ_N$ the relation
\beq
\gamma^N_{g} \ =\ \tilde c\,\one,\qquad \tilde c\in\IC^*, 
\eeq
can always be brought to the form
\beq
\hat\gamma_g^N \ =\ \one,
\eeq
defining $\hat\gamma_g = \tilde c^{-1/N}\gamma_g$. For real projective
representations, i.e.\ $\tilde c\in\IR^*$, this is only possible if $N$ is
odd or if $\tilde c>0$. If $N$ is even, two inequivalent projective 
representations arise:
\beq
\gamma_g^N \ =\ \pm\one.
\eeq
These are the cases with and without vector structure \cite{blpssw}.

If $\Gamma=\IZ_N\times\IZ_M$ (with $N,M$ both even), then four cases have
to be distinguished: $\gamma_{g_1}^N =\pm\one$, $\gamma_{g_2}^M =\pm\one$.
Only if $\gamma_{g_1}^N =\gamma_{g_2}^M=\one$ and 
$\gamma_{g_1}\gamma_{g_2}=\gamma_{g_2}\gamma_{g_1}$, does the gauge bundle 
of the orientifold have vector structure in the sense of \cite{blpssw,w}.
Thus, orientifolds with discrete torsion can never have vector structure.
However, the different boundary conditions for the gauge bundle of 
orientifolds with discrete torsion --- let us denote them by $(++)$,
$(+-)$, $(-+)$ and $(--)$ --- lead, in general, to \noneq models.
In particular, the gauge group and the spectrum are not the same in all
four cases.

As explained in appendix \ref{app_proj}, there are two types of real 
projective representations: irreducible real projective representations 
and combinations of pairs of conjugate irreducible complex representations.
Note, that if $\tilde\gamma$ is a complex projective representation of the
discrete group $\Gamma$, so is $\tilde\gamma^*=(\tilde\gamma^\top)^{-1}$.
The orientifold projection tells us that one has to take these two
representations together, whenever $\tilde\gamma\neq\tilde\gamma^*$.
In terms of matrices this means:
\beq  \label{rep2}
\gamma_g \ =\ \left(\begin{array}{cc}
                  \tilde\gamma_g & 0 \\
            0 &  c\,\delta^{-1}_g (\tilde\gamma^\top_g)^{-1} 
                 \end{array} \right) \qquad
\forall\,g\in\Gamma, 
\eeq
where $c=\pm1$ and $\delta_g$ is a phase. It can be verified that
$\gamma_g$ satisfies (\ref{gamomg_cond}) if $\gamma_\Omega$ is of the
form
\beq  \label{omega2}
\gamma (\Omega) \ =\ \left(\begin{array}{cc}
                            0 & 1 \\
                            c & 0  
                     \end{array} \right) \otimes\one_n, 
\eeq
where $n$ is the dimension of $\tilde\gamma$.
The fact that $\gamma_\Omega$ can always be chosen to be of this form
is a consequence of (\ref{gamom_cond}).

Now, let us see the restrictions that arise from the multiplication of
two group elements $g,h\in\Gamma$:
\beq
\gamma_g\gamma_h \ =\ \left(\begin{array}{cc}
                   \tilde\gamma_g \tilde\gamma_h & 0 \\
             0 &  \delta_g^{-1} \delta_h^{-1} 
                  (\tilde\gamma^\top_g)^{-1}(\tilde\gamma^\top_h)^{-1}  
                    \end{array} \right) 
\eeq
From the relation $\gamma_g\gamma_h=\alpha_{g,h}\gamma_{gh}$, eq.\
(\ref{gamma_gh}), one finds
\beq
\gamma_g\gamma_h \ =\ \alpha_{g,h}\, \left(\begin{array}{cc}
                              \tilde\gamma_{gh} & 0 \\
                0 &  \delta_g^{-1} \delta_h^{-1} \alpha_{g,h}^{-2} 
                     (\tilde\gamma^\top_{gh})^{-1}  
                                          \end{array} \right) 
                      \ =\  \alpha_{g,h}\,\gamma_{gh},
\eeq
which gives the relation between the discrete torsion $\alpha_{g,h}$
and the orientifold phase $\delta_g$: 
\be  \label{alpha_delta}
\delta_g\, \delta_h\, \alpha_{g,h}^{2} =c\,\delta_{gh}.
\ee 
This condition comes from the fact that two projective representation can be 
related if they belong to the same equivalence class of factor systems 
(see appendix \ref{app_proj} or \cite{karp,h}). The projective 
representation  $\tilde\gamma_g$ has the factor system $\alpha_{g,h}$ and 
$(\tilde\gamma_g^\top)^{-1}$  the inverse one, $\alpha_{g,h}^{-1}$. 
They are in the same equivalence class if there exist numbers $\rho_g$, 
such that $(\rho_g \rho_h/ \rho_{gh})\,\alpha_{g,h}=\alpha_{g,h}^{-1}$, 
which is just the above condition.

Interchanging $g$ and $h$ in (\ref{alpha_delta}), one finds 
$(\delta_h \delta_g/c\delta_{hg}) \alpha_{h,g} = \alpha_{h,g}^{-1}$. The 
quotient of these two relations gives the following consistency condition:
$(\alpha_{h,g}\alpha^{-1}_{g,h})^2 =1$. This tells us that the only values of 
discrete torsion compatible with the orientifold projection are 
\be  \label{beta_real}
\beta_{g,h}=\alpha_{g,h}\alpha^{-1}_{h,g}=\pm1.
\ee
This result coincides with the consistency condition that we found in the
closed string sector of orientifolds with discrete torsion.

According to the results of appendix \ref{app_proj}, the matrix $\gamma_g$ 
(whith $g=g_1^ag_2^b$) of a general real projective representation with 
discrete torsion $\eps=-1$ and with $\eta_{1/2}=0,1$, such that 
$(\gamma_{g_1})^N=(-1)^{\eta_1}\one$, $(\gamma_{g_2})^M=(-1)^{\eta_2}\one$, 
is of the form
\be \label{gamma_real}
\bigoplus_{k,l}\left((\omega_{2N}^{2k+\eta_1}\gamma_{g_1})^a\,
                    (\omega_{2M}^{2l+\eta_2}\gamma_{g_2})^b\,
            \oplus\,(\omega_{2N}^{2N-2k-\eta_1}\gamma_{g_1})^a\,
                    (\omega_{2M}^{2M-2l-\eta_2}\gamma_{g_2})^b\right)
            \otimes\one_{n_{kl}},
\ee
where 
\be
\gamma_{g_1}=\left(\ba{cc}1&0\\0&-1\ea\right),\qquad 
\gamma_{g_2}=\left(\ba{cc}0&1\\1&0\ea\right),
\ee
as given in (\ref{irrep}), and $k=0,\ldots,N/2-1$, $l=0,\ldots,M/2-1$. 
Here, we chose the factor system (\ref{orbi_fs}), which implies 
$(\gamma_{g_1})^a(\gamma_{g_2})^b=\gamma_{g_1^ag_2^b}$.

Let us now analyze some examples. Start with the $\IZ_2 \times \IZ_2$ case 
with $(++)$ boundary condition, i.e.\ $\gamma_{g_1}^2=\gamma_{g_2}^2=\one$.
Discrete torsion is possible for this model, because
$H^2(\IZ_2 \times \IZ_2,\IZ_2)=\IZ_2$. The case without discrete torsion has 
four irreducible representations, all of them are one dimensional and real. 
The case with discrete torsion $-1$ has a unique irreducible representation,
that can be taken to be real. A general representation, with $(++)$ boundary 
condition, is of the form
\beqa  \label{gam22++}
\gamma_e &= &\one_2 \otimes \one_n, \\
\gamma_{g_1} &= &\sigma_3 \otimes \one_n, \nonumber \\
\gamma_{g_2} &= &\sigma_1 \otimes \one_n, \nonumber \\
\gamma_{g_1g_2} &= &i\sigma_2 \otimes \one_n, \nonumber 
\eeqa
where $n$ is an arbitrary parameter. If the matrix $\gamma_\Omega$ is
symmetric, then it can be taken to be of the form
\be  \label{om_sym}
\gamma_\Omega=\one_2\otimes\one_n.
\ee
If it is antisymmetric, we restrict ourselves to the case of even $n$ and
take
\be \label{om_asym}
\gamma_\Omega=\one_2\otimes\left(\ba{cc}0&1\\-1&0\ea\right)\otimes\one_{n/2}.
\ee
Using (\ref{gam_omg}), with $\alpha_{\Omega,g}=\alpha_{g,\Omega}=1$, it is 
easy to check that all consistency conditions from the multiplication of 
group elements (like $g_1^2=g_2^2=e$, $\Omega g=g\Omega$, etc) are satisfied
for this choice of matrices. As we mentioned above, this choice is unique
up to equivalence.

The spectrum can be determined by taking the matrices 
$\lambda^{(0)},\lambda^{(i)}$ of the orbifold case (\ref{sol22}),
\be \label{sol_orien22}
\lambda^{(0)}\ =\ \one_2 \otimes X,\qquad 
\lambda^{(i)}\ =\ \sigma_i \otimes Z_i,
\ee
and restricting them, such that also the second condition in (\ref{ospec_cond})
is satisfied. If $\gamma_\Omega$ is symmetric, then $X$, $Z_1$, $Z_3$ are 
antisymmetric and $Z_2$ is symmetric. This corresponds to gauge group
$SO(n)$ with two adjoint fields, a traceless symmetric tensor and a singlet.
If $\gamma_\Omega$ is antisymmetric, we find\footnote{In our notation,
$USp(2k)$ is the unitary symplectic group of rank $k$.}  $USp(n)$
with two adjoint fields and an antisymmetric tensor.
The superpotential is the one of the orbifold with the above representations:
\beq  \label{sp_orien22}
W\ =\ \Tr(Z_{1}Z_{2}Z_{3}+Z_{2}Z_{1}Z_{3})
\eeq

The solution to the case with $(--)$ boundary condition is essentially
the $\IZ_2\times\IZ_2$ model of \cite{bl} without $D5$-branes.
In our conventions the $\gamma$-matrices are given by
\bea  \label{gam22--}
\gamma_e &= &\one_2 \otimes \one_{2n}, \\
\gamma_{g_1} &= &(i\sigma_1\oplus(-i\sigma_1))\otimes\one_n, \nonumber \\
\gamma_{g_2} &= &(i\sigma_3\oplus(-i\sigma_3))\otimes\one_n, \nonumber \\
\gamma_{g_1g_2} &= &i\sigma_2 \otimes \one_{2n}, \nonumber 
\eea
where we exchanged $\gamma_{g_1}$ and $\gamma_{g_2}$ with respect to
(\ref{gamma_real}) for later convenience. If $\gamma_\Omega$ is symmetric,
the spectrum consists of $USp(2n)$ gauge fields and three matter fields in 
the antisymmetric tensor representation. This is exactly the spectrum of the
99 sector of the model discussed in \cite{bl}, if $n=8$. If $\gamma_\Omega$
is antisymmetric, we find $SO(2n)$ with three symmetric tensors.

A similar analysis can be done for the boundary conditions $(+-)$ and $(-+)$. 
One finds in both cases the same spectrum as in the $(++)$ case.

To determine the spectrum of the $\IZ_2\times\IZ_4$ orientifold, it is more
convenient to use the shift formalism, as explained in appendix 
\ref{app_shift}. We take $\gamma_\Omega$ symmetric and boundary conditions
$(-+)$. Let $g_1$ resp.\ $g_2$ be the generators of $\IZ_2$ resp.\ $\IZ_4$.
We choose a basis where $\gamma_{g_2}$ is diagonal. According to appendix
\ref{app_shift}, $g_2$ is represented by the shift 
$$V_{g_2}=\frac14(0^{n_0},2^{n_0},1^{n_1},3^{n_1})$$
and $g_1$ is represented by the permutation, acting on the roots $\rho$ of the
$SO(2(n_0+n_1))$ lattice, 
$$\Pi_{g_1}:\ (\tilde\rho_1^{(n_0)},\tilde\rho_2^{(n_0)},
                 \tilde\rho_1^{(n_1)},\tilde\rho_2^{(n_1)})\ 
               \longrightarrow\ 
              (\tilde\rho_2^{(n_0)},\tilde\rho_1^{(n_0)},
                 \tilde\rho_2^{(n_1)},\tilde\rho_1^{(n_1)}),$$
where $\tilde\rho_i^{(n_0)}=(\rho_{(i-1)n_0+1},\ldots,\rho_{in_0})$
and $\tilde\rho_i^{(n_1)}=(\rho_{2n_0+(i-1)n_1+1},\ldots,\rho_{2n_0+in_1})$.
As explained in appendix \ref{app_shift} no shift associated to $g_1$ is
needed, whereas in the $(++)$ case we would need an additional shift
$V_{g_1}=\qt(1^{\tilde n_0})$.
(Note that $g_1$ and $g_2$ are exchanged with respect to the notation in
appendix \ref{app_shift}). The gauge group is determined by finding all
linear combinations of roots $\rho$ that satisfy $\rho\cdot V_{g_2}=0$
and $\Pi_{g_1}(\rho)=\rho$. They are of the form
\newpage
\begin{eqnarray*}
  (\underline{+,+,0^{n_0-2}},0^{n_0},0^{2n_1})+
  (0^{n_0},\underline{+,+,0^{n_0-2}},0^{2n_1}),\\
  (\underline{-,-,0^{n_0-2}},0^{n_0},0^{2n_1})+
  (0^{n_0},\underline{-,-,0^{n_0-2}},0^{2n_1}),\\
  (\underline{+,-,0^{n_0-2}},0^{n_0},0^{2n_1})+
  (0^{n_0},\underline{+,-,0^{n_0-2}},0^{2n_1}),\\
  (0^{2n_0},\underline{+,0^{n_1-1}},\underline{0^{n_1-1},+})+
  (0^{2n_0},\underline{0^{n_1-1},+},\underline{+,0^{n_1-1}}),\\
  (0^{2n_0},\underline{-,0^{n_1-1}},\underline{0^{n_1-1},-})+
  (0^{2n_0},\underline{0^{n_1-1},-},\underline{-,0^{n_1-1}}),\\
  (0^{2n_0},\underline{+,-,0^{n_1-2}},0^{n_1})+
  (0^{2n_0},0^{n_1},\underline{+,-,0^{n_1-2}}).
\end{eqnarray*}
Underlining means that one has to take all correlated permutations of the
two terms in each line, such that they are invariant under $\Pi_{g_1}$.
The first three lines give $n_0(2n_0-1)-n_0$ roots, which, together with
$n_0$ Cartan generators, form the gauge group $SO(2n_0)$. The last three
lines give $n_1(2n_1+1)-n_1$ roots, which, together with $n_1$ Cartan
generators, form the gauge group $USp(2n_1)$. The matter fields from the
first complex plane are obtained from the roots that satisfy 
$\rho\cdot V_{g_2}=0$ and $\Pi_{g_1}(\rho)=-\rho$. These are just the
antisymmetric combinations of the roots above that formed the gauge
group. Thus, we find an adjoint field of $SO(2n_0)$ and an antisymmetric
tensor of $USp(2n_1)$. For the second complex plane, one has the condition
$\rho\cdot V_{g_2}=1/4$ and $\Pi_{g_1}(\rho)=-\rho$. The corresponding roots
are
\begin{eqnarray*}
  (\underline{+,0^{n_0-1}},0^{n_0},\underline{+,0^{n_1-1}},0^{n_1})-
  (0^{n_0},\underline{+,0^{n_0-1}},0^{n_1},\underline{+,0^{n_1-1}}),\\
  (0^{n_0},\underline{-,0^{n_0-1}},\underline{-,0^{n_1-1}},0^{n_1})-
  (\underline{-,0^{n_0-1}},0^{n_0},0^{n_1},\underline{-,0^{n_1-1}}),\\
  (\underline{-,0^{n_0-1}},0^{n_0},\underline{+,0^{n_1-1}},0^{n_1})-
  (0^{n_0},\underline{-,0^{n_0-1}},0^{n_1},\underline{+,0^{n_1-1}}),\\
  (\underline{+,0^{n_0-1}},0^{n_0},0^{n_1},\underline{-,0^{n_1-1}})-
  (0^{n_0},\underline{+,0^{n_0-1}},\underline{-,0^{n_1-1}},0^{n_1}).
\end{eqnarray*}
They form a matter field transforming in the bifundamental $(\fund,\fund)$ 
representation of the gauge group. The matter fields from the third complex
plane correspond to roots that satisfy $\rho\cdot V_{g_2}=-1/4$ and 
$\Pi_{g_1}(\rho)=\rho$. Again, one finds $4n_0n_1$ roots, giving a second
bifundamental.

A similar analysis can be performed for the other boundary conditions,
$(+-)$, $(++)$, $(--)$. The result is shown in tables \ref{noncomorient++}%
--\ref{noncomorient--}. One finds that the $(+-)$ model and the $(--)$
model have \nonab gauge anomalies in the 33 sector. In the language of
the previous subsection, these two boundary conditions correspond to 
${\mu_1}^3=-1$. There, we saw that precisely these models need $D7$-branes
to cancel the tadpoles. A minimal choice consists
in a set of $D7_2$ branes satisfying 
$\Tr\gamma_{(0,2),7_2}=-16\,{\mu_3}^3\delta_{(0,1)}$.
Let us consider the $(--)$ model with $\gamma_\Omega$ symmetric. According
to (\ref{orien_fs}), this gives $\delta_g=1$. From (\ref{vec_cond2}), 
we have ${\mu_1}^{7_2}=-{\mu_1}^3$ and ${\mu_3}^{7_2}=-{\mu_3}^3$. Thus the
theory on the $D7_2$-branes has $(++)$ boundary condition. As explained in
\cite{gp,bl}, $\gamma_{\Omega_{7_2}}$ has to be antisymmetric.\footnote{If,
instead of the standard $\Omega$-projection discussed by Gimon and Polchinski
\cite{gp}, we use the alternative projection proposed by Dabholkar and Park
in $D=6$ \cite{dp}, $\gamma_{\Omega,7_2}$ and $\gamma_{\Omega,3}$ have 
the same symmetry.}
In table \ref{noncomorient++} the general solution for the gauge theory on 
a set of $D$-branes at a $\IZ_2\times\IZ_4$ singularity is shown for the 
case of symmetric $\gamma_\Omega$. Knowing that changing $\gamma_\Omega$ 
from symmetric to antisymmetric exchanges $SO$-factors with $USp$-factors, 
we find that the gauge theory on the $D7_2$-branes is 
$USp(2m_0)\times USp(2m_1)$.
In order to cancel the tadpoles, we need $2(2m_0-2m_1)=16$. The $37_2$ sector 
gives matter fields transforming in the representations $(\fund;\fund,\one)$
and $(\antifund;\one,\fund)$ under the total gauge group 
$U(n_0+n_1)\times USp(2m_0)\times USp(2m_1)$. The total anomaly is thus given
by $2m_0-2m_1-8$. We see that the condition of anomaly freedom is equivalent
to the condition of vanishing tadpoles.
 
In the same way the $\IZ_2\times\IZ_6$ and the $\IZ_2\times\IZ_6'$ 
orientifold can be constructed. The result is shown in tables
\ref{noncomorient++}--\ref{noncomorient--}. Again, some of these
models have \nonab gauge anomalies in the 33 sector. It can be seen
that the same models require $D7$-branes for tadpole cancellation.

\begin{table}[ht!]
\renewcommand{\arraystretch}{1.25}
$$\begin{array}{|c|c|c|} \hline
\Gamma &\rm gauge\ group &\rm matter\quad \hbox{\small (33 sector)} \\
\hline
\IZ_2 \times \IZ_2 & SO(2n) &  2\,adj + \Ysymm \\
\hline
\IZ_2 \times \IZ_4 & SO(2n_0) \times SO(2n_1) 
                   & 2\,(\fund,\fund) +(\one,\Ysymm) +(\Yasymm,\one) \\
\hline
\IZ_2 \times \IZ_6 & SO(2n_0) \times U(n_1+n_2)
                   &(\fund,\antifund) +(\fund, \fund)\\ 
                  &&+(\one,\bYasymm) +(\one,\Ysymm) + (adj,\one) +(\one,adj)\\
\hline
\IZ_2 \times \IZ_6' & SO(2n_0) \times U(n_1+n_2) 
                    &3\,(\fund, \fund) + 2(\one,\bYasymm) + (\one,\bYsymm)\\
\hline
\end{array}$$
\caption{Open string spectrum for some non-compact orientifolds with
discrete torsion $\eps=-1$ and boundary condition $(++)$, i.e.\
$\gamma_{g_1}^N=\one$, $\gamma_{g_2}^M=\one$.
\label{noncomorient++}}
\vskip10mm
\end{table}

\vspace{-1cm}
\begin{table}[ht!]
\renewcommand{\arraystretch}{1.25}
$$\begin{array}{|c|c|c|} \hline
\Gamma &\rm gauge\ group &\rm matter\quad \hbox{\small (33 sector)} \\
\hline
\IZ_2 \times \IZ_2 & SO(2n)  &2\,adj + \Ysymm \\
\hline
\IZ_2 \times \IZ_4 &U(n_0+n_1) & adj + \Yasymm + 2\,\bYasymm + \Ysymm  \\
\hline
\IZ_2 \times \IZ_6  & SO(2n_1) \times U(n_0+n_2)
                    &(\fund,\antifund) +(\fund, \fund)\\ 
                   &&+(\one,\bYasymm)+(\one,\Yasymm)+(\Ysymm,\one)+(\one,adj)\\
\hline
\IZ_2 \times \IZ_6' & SO(2n_1) \times U(n_0+n_2) 
                    & 3\,(\fund, \antifund)+2\,(\one,\Yasymm)+(\one,\Ysymm)\\
\hline
\end{array}$$
\caption{Open string spectrum for some non-compact orientifolds with
discrete torsion $\eps=-1$ and boundary condition $(+-)$, i.e.\
$\gamma_{g_1}^N=\one$, $\gamma_{g_2}^M=-\one$.
\label{noncomorient+-}}
\end{table}

\begin{table}[ht!]
\renewcommand{\arraystretch}{1.25}
$$\begin{array}{|c|c|c|} \hline
\Gamma &\rm gauge\ group &\rm matter\quad \hbox{\small (33 sector)} \\
\hline
\IZ_2 \times \IZ_2 & SO(2n) &2\ adj + \Ysymm\\
\hline
\IZ_2 \times \IZ_4 & SO(2n_0) \times USp(2n_1) 
                   & (adj,\one) + (\one,\Yasymm) + 2\,(\fund,\fund) \\
\hline
\IZ_2 \times \IZ_6 & SO(2n_0) \times U(n_1+n_2)
                   & (\fund,\antifund)+(\fund,\fund)+(\one,\Yasymm)
                     +(\one,\bYsymm)\\ 
                  && +(adj,\one)+(\one,adj) \\
\hline
\IZ_2 \times \IZ_6' & SO(2n_0) \times U(n_1+n_2) 
                    & 3\,(\fund,\fund)+2,(\one,\bYasymm)+(\one,\bYsymm)\\
\hline
\end{array}$$
\caption{Open string spectrum for some non-compact orientifolds with 
discrete torsion $\eps=-1$ and boundary condition $(-+)$, i.e.\
$\gamma_{g_1}^N=-\one$, $\gamma_{g_2}^M=\one$.
\label{noncomorient-+}}
\end{table}

\begin{table}[ht!]
\renewcommand{\arraystretch}{1.25}
$$\begin{array}{|c|c|c|} \hline
\Gamma &\rm gauge\ group &\rm matter\quad \hbox{\small (33 sector)} \\
\hline
\IZ_2 \times \IZ_2 & USp(2n) &3\,\Yasymm \\
\hline
\IZ_2 \times \IZ_4 & U(n_0+n_1) & adj+2\,\Yasymm+\bYasymm+\bYsymm \\
\hline
\IZ_2 \times \IZ_6 & USp(2n_1) \times U(n_0+n_2)
                   & (\fund,\antifund) +(\fund, \fund)\\ 
                  && +(\one,\bYasymm)+(\one,\Yasymm)+(\Yasymm,\one)
                     +(\one,adj) \\
\hline
\IZ_2 \times \IZ_6' & USp(2n_1) \times U(n_0+n_2) 
                    & 3\,(\fund, \antifund) + 3\,(\one,\Yasymm)\\
\hline
\end{array}$$
\caption{Open string spectrum for some non-compact orientifolds with
discrete torsion $\eps=-1$ and boundary condition $(--)$, i.e.\
$\gamma_{g_1}^N=-\one$, $\gamma_{g_2}^M=-\one$.
\label{noncomorient--}}
\vskip10mm
\end{table}

In general, one finds the following gauge groups for $\IZ_2\times\IZ_N$
orientifolds with discrete torsion (if $\gamma_\Omega$ is symmetric):

$\bullet$ boundary condition $(++)$:
\be  \label{gaugegr-+} \ba{ll}
N=4k+2: &G=SO(2n_0)\times\prod_{i=1}^kU(n_i+n_{N/2-i}),\\
N=4k:   &G=SO(2n_0)\times SO(2n_k)\times\prod_{i=1}^{k-1}U(n_i+n_{N/2-i}).
\ea\ee

$\bullet$ boundary condition $(+-)$:
\be  \label{gaugegr--} \hspace*{-1.5cm}\ba{ll}
N=4k+2: &G=SO(2n_k)\times\prod_{i=0}^{k-1}U(n_i+n_{N/2-1-i}),\\
N=4k:   &G=\prod_{i=0}^{k-1}U(n_i+n_{N/2-1-i}).
\ea\ee

$\bullet$ boundary condition $(-+)$:
\be  \label{gaugegr++} \ba{ll}
N=4k+2: &G=SO(2n_0)\times\prod_{i=1}^kU(n_i+n_{N/2-i}),\\
N=4k:   &G=SO(2n_0)\times USp(2n_k)\times\prod_{i=1}^{k-1}U(n_i+n_{N/2-i}).
\ea\ee

$\bullet$ boundary condition $(--)$:
\be  \label{gaugegr+-} \hspace*{-1.5cm}\ba{ll}
N=4k+2: &G=USp(2n_k)\times\prod_{i=0}^{k-1}U(n_i+n_{N/2-1-i}),\\
N=4k:   &G=\prod_{i=0}^{k-1}U(n_i+n_{N/2-1-i}).
\ea\ee

\subsection{Resolution of the singularities} 

Some of the deformations that are available in the orbifold case are absent 
in the orientifold case.

Take the easiest example: $\IZ_2\times\IZ_2$. In the orbifold case one can 
find three possible deformations \cite{d}:
\beq 
\Delta W\ =\ \sum_{i=1}^{3} \zeta_i \Tr Z_i.
\label{defo}
\eeq

For the $\IZ_2\times\IZ_2$ with $(++)$ boundary condition, only one of the 
deformations survives the orientifold projection. Because of the symmetry 
of the $Z_1$, $Z_3$ (they are antisymmetric) and $Z_2$ (symmetric) matrices,
one has:
\beq
\zeta_{1} \Tr Z_1\ =\ 0,\qquad \zeta_{3} \Tr Z_3\ =\ 0. 
\label{defnc}
\eeq

This indicates that the orientifold can only be deformed to the $\IZ_2$ 
singularity. The additional deformation leading to the conifold is frozen in
the orientifold case. Something similar happened in the $\IZ_2 \times \IZ_2$ 
orientifold without discrete torsion \cite{pru}.

\section{Conclusions}

We have seen that only real values of the discrete torsion parameter $\eps$
are allowed for orientifold models, in contrast to the orbifold case, where
$\eps$ can take complex values. This condition can be understood from several
viewpoints. In the open string sector this is related to the fact that the
$\gamma$-matrices form {\em real} projective representations. These are
classified by $H^2(\Gamma,\IR^\ast)= H^2(\Gamma,\IZ_2)$, in a similar manner
as complex projective representations are characterized by 
$H^2(\Gamma,\IC^\ast)= H^2(\Gamma, U(1))$. As a consequence, only $\eps=\pm1$
is allowed for orientifolds. In the closed string spectrum, the condition
of real $\eps$ is related to the matching between left and right moving
degrees of freedom. In general, this matching is impossible if the Hodge 
numbers $h^{1,2}$ and  $h^{2,1}$ from one twisted sector are different.
One finds that $h^{1,2}=h^{2,1}$ is only guaranteed in each twisted sector
if $\eps=\pm1$.
Finally, one finds an inconsistency in the tadpole cancellation conditions
for non-real $\eps$. The characters of the projective representation $\gamma$
must have a precise value to cancel the  Klein bottle contribution.
This conditions cannot be satisfied if arbitrary values of the discrete 
torsion are allowed.

For the \ZNM\ orientifold \nontriv discrete torsion is only possible if
$N$ and $M$ are both even. In this case, there are four \noneq orientifold
models. This classification is based on the possibility of imposing
different boundary conditions (`vector structures') on the $\gamma$-matrices. 
The tadpole conditions are different for each case. 
Some of them require $D7$-branes for consistency. This condition is equivalent
to the requirement that \nonab gauge anomalies be absent.

We have also analyzed the resolution of the $\IZ_2\times\IZ_2$ singularity
in both cases, orbifold and orientifold. In the orbifold case, one can get
a $\IZ_2$ singularity and a conifold by the deformations of the F-flatness
conditions. In the orientifold only the $\IZ_2$ singularity can be obtained.

This discussion can be generalized to compact orientifolds with discrete 
torsion \cite{kr}. The S-dual heterotic models are related to higher level 
Kac-Moody algebras. In particular, as the only \nontriv discrete torsion is
$\eps=-1$, we will find heterotic duals realized at Kac-Moody level 2 
\cite{afiu,afiuv,fiq2}.

\vskip3cm

\centerline{\bf Acknowledgements}
It is a pleasure to thank Angel Uranga and Luis~Ib\'a\~nez for many 
helpful ideas and comments on the manuscript. 
The work of M.K.\ is supported by a TMR network of the European Union, 
ref. FMRX-CT96-0090. The work of R.R.\ is supported by the MEC through a FPU Grant.

\newpage
\begin{center}\huge\bf Appendix \end{center}
\begin{appendix}

\section{Projective representations} \label{app_proj}
The matrices that represent the action of the orbifold group on the Chan-Paton
indices of open strings form projective representations. In this appendix, we
review some useful facts about the latter (for an introduction to this subject
see \cite{karp,h}).
Let $\Gamma$ be a finite group. A projective representation is a mapping
$\Gamma\ \to \ GL(n,\IK)$, which associates to each element $g\in\Gamma$
a matrix $\gamma_g\in GL(n,\IK)$ that satisfies the conditions\footnote{%
Equivalently a projective representations can be defined as a homomorphism 
$\Gamma\ \to\ PGL(n,\IK)$.}
\be \label{proj_rep}
\gamma_e=\one,\qquad \gamma_g\gamma_h=\alpha_{g,h}\gamma_{gh}\qquad\forall
g,h\in\Gamma,
\ee
where $e$ is the neutral element of $\Gamma$ and $\alpha_{g,h}\in\IK^\ast$ 
are arbitrary \nonze numbers. These numbers are called the {\em factor system}
of the projective representation $\gamma$.
Using associativity, one immediately obtains
\be \label{cocycle}
 \alpha_{g,e}=\alpha_{e,g}=1,\qquad \alpha_{g,hk}\alpha_{h,k}=
 \alpha_{g,h}\alpha_{gh,k}\qquad\forall g,h,k\in\Gamma.
\ee
A map $\alpha:\ \Gamma\times\Gamma\ \to\ \IK^\ast$,\ \ 
$(g,h)\ \mapsto\ \alpha_{g,h}$, with the above properties is
called a {\em cocycle}. The set of all cocycles is denoted by
$Z^2(\Gamma,\IK^\ast)$. A projective representation $\hat\gamma$ is
considered equivalent to $\gamma$ if it is obtained by substituting
$\hat\gamma_g=\rho_g\gamma_g$, with $\rho_g\in\IK^\ast$. The cocycles are
then related by
\be \label{equiv_cocycle}
\hat\alpha_{g,h}=\alpha_{g,h}\,\rho_g\rho_h\rho_{gh}^{-1}.
\ee
This motivates the following definition. Given a map $\rho:\ \Gamma\ 
\to\ \IK^\ast$, $g\ \mapsto\ \rho_g$, with $\rho_e=1$, one defines the 
associated {\em coboundary} by 
$\delta\rho:\ \Gamma\times\Gamma\ \to\ \IK^\ast$,
$(g,h)\ \mapsto\ \rho_g\rho_h\rho_{gh}^{-1}$. The set of all coboundaries,
$B^2(\Gamma,\IK^\ast)$, is a subset of $Z^2(\Gamma,\IK^\ast)$. Two
cocycles are equivalent if they differ only by a coboundary. We see that
the \noneq projective representations are characterized by the elements
of $H^2(\Gamma,\IK^\ast)=Z^2(\Gamma,\IK^\ast)/B^2(\Gamma,\IK^\ast)$.

If $\Gamma$ is Abelian, one finds from (\ref{proj_rep}) that
\be \label{def_beta}
\gamma_g\gamma_h=\beta_{g,h}\gamma_h\gamma_g,\qquad
{\rm where\ } \beta_{g,h}=\alpha_{g,h}\alpha_{h,g}^{-1}.
\ee
The $\beta_{g,h}$ only depend on the equivalence class of the $\alpha_{g,h}$.
Furthermore they satisfy
\be \label{beta_cond}
\beta_{g,g}=1,\qquad \beta_{g,h}=\beta_{h,g}^{-1},\qquad
\beta_{g,hk}=\beta_{g,h}\beta_{g,k}\qquad \forall g,h,k\in\Gamma.
\ee
It is clear that to each element $[\alpha]$ of $H^2(\Gamma,\IK^\ast)$ there 
corresponds a unique cocycle\footnote{Note that $\beta$ is a cocycle because 
it satisfies (\ref{cocycle}), but it is not a factor system because its 
definition differs from (\ref{proj_rep}).} $\beta$, as given in 
(\ref{def_beta}). On the other hand to each $\beta$ there corresponds a unique 
(up to equivalence) factor system $\alpha_{g,h}=\sqrt{\beta_{g,h}}$. As a 
consequence, the projective representations of an Abelian finite group 
$\Gamma$ can be characterized either by (\ref{proj_rep}), (\ref{cocycle}) or 
by the first eq.\ of (\ref{def_beta}) and the three eqs.\ of (\ref{beta_cond}).
Both descriptions are equivalent up to a transformation (\ref{equiv_cocycle}).
Sometimes it is useful to fix as many of the $\alpha_{g,h}$ as possible
without putting any restriction on the equivalence class $[\alpha]$.
A convenient choice is:
\be  \label{orbi_fs}
\alpha_{g_1^a,g_2^b}=\alpha_{g_1,g_1^a}=\alpha_{g_2,g_2^b}=1, \qquad
a=1,\ldots,N,\quad b=1,\ldots,M.
\ee
This corresponds to choosing a set of matrices that satisfies
\be  \label{gam_choice}
(\gamma_{g_1})^a(\gamma_{g_2})^b=\gamma_{g_1^ag_2^b}.
\ee

For physical reasons we restrict ourselves to unitary projective
representations over the complex or real numbers, 
i.e.\ $\gamma_g\gamma_g^\dagger=\one$ and $\IK=\IC$ or $\IK=\IR$. 
One can prove that\footnote{For the cocycles related to unitary projective 
representations this is obvious. In general, this is a consequence of 
proposition 2.3.10 and lemma 2.3.19 of \cite{karp}.} 
\be \label{Htwo_rels}
H^2(\Gamma,\IC^\ast)=H^2(\Gamma,U(1)),\qquad
H^2(\Gamma,\IR^\ast)=H^2(\Gamma,\IZ_2).
\ee

In this article we will be mainly interested in the case where
$\Gamma=\IZ_N\times\IZ_M$. Let $g_1$ and $g_2$ be the generators 
of $\IZ_N$ and $\IZ_M$ respectively and denote the element 
$g_1^ag_2^b\in\Gamma$ by $(a,b)$. The form of the $\beta$ cocycles
is completely fixed by the conditions (\ref{beta_cond}).
If $\IK=\IC$, there are $p=\gcd(N,M)$ \noneq cocycles, given by
\be \label{beta}
\beta_{(a,b),(a^\prime,b^\prime)}=\omega_p^{m(ab^\prime-ba^\prime)},
\qquad {\rm with\ }\omega_p=e^{2\pi i/p},\ m=1,\ldots,p.
\ee
The different projective representations are therefore determined by 
the parameter $\eps=\omega_p^m$. Choosing the factor system (\ref{orbi_fs}),
one finds a simple relation for the product of two $\gamma$-matrices:
\be  \label{gam_prod}
\gamma_{(a,b)}\gamma_{(c,d)}=\eps^{-bc}\gamma_{(a+c,b+d)}.
\ee

The possibility of having $\eps\neq1$
is the open string analogue of the discrete torsion in closed string 
orbifolds discussed in \cite{v,fiq,vw}. The open string models with
discrete torsion are distinguished by the integer $s=p/\gcd(m,p)$ 
(this is the smallest \nonze number such that $\eps^s=1$). 
According to \cite{beren}, any irreducible projective representation 
$\cR^{(s)}_{\rm irr}$ is $s$-dimensional and (up to projective equivalence)
of the form
\be  \label{irrep}
\gamma_{g_1}\ =\ \diag(1,\eps^{-1},\eps^{-2},...,\eps^{-(s-1)}),\qquad
\gamma_{g_2}\ =\ 
\left(\ba{ccccc}
0 & 1 & 0 & \dots &0\\
0 & 0 & 1 & \dots & \\
\vdots & & \ddots & \ddots \\
0 &   & \dots & 0 &1\\
1 & 0 & \dots &  &0
\ea\right).
\ee
All the the irreducible projective representations $\cR^{(s)}_{{\rm irr,}k,l}$ 
that are {\em linearly} \noneq (i.e.\ they belong to different factor systems
but, of course, are projectively equivalent) can be obtained by multiplying 
the matrices in (\ref{irrep}) by phases:
\be \label{irreplin}
\hat\gamma_{g_1}=\omega_N^k\gamma_{g_1},\quad k=0,\ldots,{N\over s}-1,\qquad
\hat\gamma_{g_2}=\omega_M^l\gamma_{g_1},\quad l=0,\ldots,{M\over s}-1.
\ee
A general projective representation $\cR^{(s)}$ is a direct sum of
irreducible blocks $\cR^{(s)}_{{\rm irr,}k,l}$:
\be \label{def_Rs}
\cR^{(s)}=\bigoplus_{k,l}n_{kl}\cR^{(s)}_{{\rm irr,}k,l}.
\ee
This representation has dimension $s\sum_{k,l}n_{kl}$. The regular
represention, of dimension $|\Gamma|=NM$, is obtained by setting
$n_{kl}=s\ \forall\, k,l$. Let $\gamma^{(s)}_g$ denote the matrix
associated to the group element $g$ in the general projective
representation $\cR^{(s)}$. The {\em character} of $g$ in $\cR^{(s)}$
is defined to be the trace of this matrix. Denoting again $g=g_1^ag_2^b$
by $(a,b)$ we find
\be \label{com_char}
\chi^{(\cR^{(s)})}(g)\equiv\Tr\gamma^{(s)}_g=\left\{\ba{ll}
  0 &{\rm if}\ (a,b)\not\in s\IZ\times s\IZ\\
  s\sum_{k,l}n_{k,l}\omega_N^{ka}\omega_M^{lb} &{\rm if}\ 
    (a,b)\in s\IZ\times s\IZ \ea\right.
\ee

The solutions for $\IK=\IR$ can be obtained by restricting $\beta$ to 
be real. The \noneq cocycles correspond to setting $m=p/2$ (if $p$ is
even) or $m=p$ in (\ref{beta}). We see that discrete torsion is only 
possible if $p$ is even. The only \nontriv cocycle in this case is 
$\beta_{(a,b),(a^\prime,b^\prime)}=(-1)^{ab^\prime-ba^\prime}$.
The irreducible projective representations are again of the form
(\ref{irrep}), with $\eps=-1$ and $s=2$. However, there are four
different complex projective representations which are not equivalent
over the real numbers and which have a real factor system\footnote{We used
that $M$ and $N$ are even, which is true because $\gcd(M,N)$ is even.}:
\bea  \label{vecstruc}
&&{\rm (i)}\quad \gamma_{g_1},\quad \gamma_{g_2}, \nonumber\\
&&{\rm (ii)}\quad \omega_{2N}\gamma_{g_1},\quad \gamma_{g_2}, \\
&&{\rm (iii)}\quad \gamma_{g_1},\quad \omega_{2M}\gamma_{g_2}, \nonumber\\
&&{\rm (iv)}\quad \omega_{2N}\gamma_{g_1},\quad \omega_{2M}\gamma_{g_2}, 
\nonumber
\eea
where $\gamma_{g_{1/2}}$ are given in (\ref{irrep}). In general, an 
irreducible complex projective representation is of the form 
$\tilde\gamma_{g_{1/2}}=\rho_{1/2}\gamma_{g_{1/2}}$, with $\rho_{1/2}\in\IC^*$.
The condition that the factor system be real implies 
$(\tilde\gamma_{g_1})^N=c_1\,\one$, $(\tilde\gamma_{g_2})^M=c_2\,\one$,
with $c_{1/2}\in\IR^*$. Therefore, $\rho_1^N=\pm1$ and $\rho_2^M=\pm1$. These
are the four possibilities of (\ref{vecstruc}). Defining 
$\eta=(\eta_1,\eta_2)$, with $\eta_i\in\{0,1\}$ by
\be  \label{eta}
(\gamma_{g_1})^N=(-1)^{\eta_1}\one,\qquad (\gamma_{g_2})^M=(-1)^{\eta_2}\one,
\ee
we can give the form of the linearly \noneq real projective representations:
\be \label{real_irrep}
\cR^{\rm real}_{{\rm irr,}k,l,\eta}=\left\{\ba{ll}
\cR^{(2)}_{{\rm irr,}k,l,\eta} &{\rm if}\ k=l=\eta_1=\eta_2=0\\
\cR^{(2)}_{{\rm irr,}k,l,\eta}\oplus(\cR^{(2)}_{{\rm irr,}k,l,\eta})^\ast 
                              &{\rm else}
\ea\right.,
\ee
where $\cR^{(2)}_{{\rm irr,}k,l}$ is obtained from (\ref{irrep}) by multiplying
$\gamma_{g_1}$ with phases $\omega_N^k\omega_{2N}^{\eta_1}$ and
$\gamma_{g_2}$ with phases $\omega_M^l\omega_{2M}^{\eta_2}$ as in 
(\ref{irreplin}) and (\ref{vecstruc}). These projective representations are
either two-dimensional (first line of (\ref{real_irrep})) or
four-dimensional (second line of (\ref{real_irrep})). Again, a general 
projective representation $\cR^{\rm real}_\eta$  is a direct sum of 
irreducible blocks $\cR^{\rm real}_{{\rm irr,}k,l,\eta}$:
\be \label{def_Rreal}
\cR^{\rm real}_\eta=\bigoplus_{k,l}n_{kl}\cR^{\rm real}_{{\rm irr,}k,l,\eta}\,.
\ee
This representation has dimension $2rn_{0,0}+4\sum_{(k,l)\neq(0,0)}n_{kl}$,
where $r=2^{|\eta_1-\eta_2+\eta_1\eta_2|}$.
The regular representation is obtained by setting $n_{0,0}=2/r$ and
$n_{kl}=1\ \forall\, (k,l)\neq(0,0)$. For the character of an element
$g=g_1^ag_2^b$ we find
\bea \label{real_char}
\chi^{(\cR^{\rm real}_\eta)}(g) &\equiv &\Tr\gamma^{\rm real}_g\\
  &= &\left\{\ba{ll}
     0 &\hbox{if $a$ or $b$ is odd}\\
     2\,r\,n_{0,0}+4\sum_{(k,l)\neq(0,0)}n_{k,l}
     \R(\omega_N^{(k+\eta_1/2)a}\omega_M^{(l+\eta_2/2)b}) 
    &\hbox{if $a$ and $b$ are even}\\
    \hbox{\footnotesize with $r=1$ if $\eta_1=\eta_2=0$ and $r=2$ else}
   \ea\right. \nonumber
\eea

\section{Tadpoles for $\bZ_N\times\bZ_M$ orientifolds}
\label{app_tadpole}

We sketch the calculation of the tadpoles for non-compact 
$\IZ_N \times \IZ_M$ orientifolds. Basically, we follow the
work of \cite{z,pru,afiv}.

The set-up consists of several $D3$-branes at the orbifold singularity plus 
some $D7_i$-branes, where the $i$ denotes the complex plane with 
Dirichlet conditions. We take the orientifold involution 
$\Omega' = \Omega (-1)^{F_L}R_1 R_2 R_3$ (other types can be considered, 
as in \cite{pru}).

The procedure we will follow is based on the computation of the tadpoles for 
$T^6 / (\IZ_N \times \IZ_M)$ and then taking the non-compact limit. In order 
to take this limit, one must ignore the untwisted tadpoles and the 
contributions with an inversely proportional dependence on the volume. 
The dependence proportional to the volume is related to some factors from 
the momentum modes which become continuous in the non-compact limit.

To simplify the notation, let us define \cite{pru} 
$s_i = \sin(\pi\bar k\cdot\bar v_i)$, 
$c_i = \cos(\pi\bar k\cdot\bar v_i)$ and 
$\tilde s_i = \sin(2\pi\bar k\cdot\bar v_i)$. 
We will not give explicitly the volume dependence, but this dependence can be 
extracted from the zeros and divergences of each contribution. 

The cylinder amplitude can be split into four sectors: $33$, 
$7_i7_i$, $7_i7_j$ and $7_i 3$. 
All the contributions can be neatly recast:
\beq  \label{tadcyl}
\cC = \sum_{\bar k=(1,1)}^{(N,M)} \frac{1}{8s_1s_2s_3} 
        \left[8s_1s_2s_3 \Tr\gamma_{\bar k,3} 
     +{\textstyle \sum^3_{i=1}}2s_i\Tr\gamma_{\bar k,7_i}\right]^2.
\eeq

Note that in the orbifold case, this is the only contribution, that has
to be considered.

Because of the $J$ operation that relates the $\bar{k}$ with the
$-\bar{k}$ twisted sector, the contributions to the Klein bottle come from the
untwisted and order-two twisted sectors. Three cases must be distinguished:
\begin{itemize}
\item[a)] $N$ and $M$ are both odd: There are no order-two twisted sectors.

\item[b)] $N$ is odd and $M$ is even: There is only one order-two sector:
$\bar k_1=(0,M/2)$. It fixes the first complex plane. Note that in this case 
no discrete torsion is allowed. The Klein bottle contribution is of the form:
\beq
\cK = \sum_{\bar k}\left( \cK_0 (\bar{k}) + \cK_1 (\bar{k}) \right),
\eeq
where $\cK_0$ is the contribution from the untwisted sector and  $\cK_1$ is 
the contribution from the order-two twisted sector that fixes the first 
complex plane.

\item[c)] $N$ and $M$ are both even: There are three order-two sectors: 
$\bar k_3=(N/2,0)$, $\bar k_1=(0,M/2)$ and $\bar k_2=(N/2,M/2)$. 
The Klein bottle contribution can be written as
\beq
\cK = \sum_{\bar k}\left( \cK_0(\bar{k})+\cK_1(\bar{k})
                    +\cK_2(\bar{k})+\cK_3(\bar{k}) \right),
\eeq
where $\cK_i$ is the contribution from the order-two twist that fixes the
$i$-th complex plane. The untwisted sector contribution is of the form
\beq
\cK_0 (\bar{k}) = 16 \frac{2\tilde{s}_1 2\tilde{s}_2 2\tilde{s}_3}
                          {4c^2_1 4c^2_2 4c^2_3},
\eeq
and the order-two twisted contributions are
\beq
\cK_i (\bar{k}) = - 16 \tilde\eps_i \beta_i \frac{2\tilde{s}_i}{4c^2_i},
\eeq
where $\tilde\eps_i$ is a sign that weights the contribution of the 
sector that fixes the $i$-th complex plane and 
$\beta_i= \beta_{\bar k_i,\bar k}$.
\end{itemize}

This Klein bottle contributions can be rewritten for each of the three cases:
\begin{itemize}
\item[a)] $N$ odd, $M$ odd:
\beq
\cK =\sum_{\bar k=(1,1)}^{(N,M)} \frac{1}{8\tilde{s}_1\tilde{s}_2\tilde{s}_3} 
                   [32 s_1s_2s_3]^2.
\eeq

\item[b)] $N$ odd, $M$ even:
\beq
\cK =\sum_{\bar k=(1,1)}^{(N,M/2)}\frac{1}{8\tilde{s}_1\tilde{s}_2\tilde{s}_3} 
                      [32 (s_1s_2s_3 - \tilde\eps_1 s_1 c_2 c_3)]^2.
\eeq

\item[c)] $N$ even, $M$ even:
\beq
\cK =\sum_{\bar k=(1,1)}^{(N/2,M/2)}\frac{1}{8\tilde{s}_1\tilde{s}_2
            \tilde{s}_3} 
    [32 (s_1s_2s_3 - \sum_{i\neq j\neq k} \tilde\eps_i\beta_i s_i c_j c_k)]^2.
\eeq
\end{itemize} 

Finally, one must consider the M\"obius strip contribution. For each of the 
four types of branes there is a contribution:
\beq
\cM_3 (\bar{k})  =  - 8\cdot8s_1s_2s_3 
   \Tr\left(\gamma^{-1}_{\Omega\bar{k},3}\gamma^\top_{\Omega\bar{k},3}\right),
\eeq
\beq
\cM_{7_i} (\bar{k})  =  8 \frac{8s_i c_j c_k}
                                    {4c^2_j 4c^2_k} 
         \Tr\left(\gamma^{-1}_{\Omega\bar{k},{7_i}}
              \gamma^\top_{\Omega\bar{k},{7_i}}\right).
\eeq

Using the properties (\ref{gamomg_cond}), (\ref{gam_omg}), (\ref{gam_prod}) 
of projective representations, we find 
\beq
\gamma_{\Omega\bar k}^{-1}\gamma_{\Omega\bar k}^\top
=\delta_{\bar k} (\gamma_{\bar k}^2)^\top
=\eps^{-k_1k_2}\delta_{\bar k}(\gamma_{\bar 2k})^\top,
\eeq
where $\bar k=(k_1,k_2)$ and we chose a factor system such that
$(\gamma_{(1,0)})^{k_1}(\gamma_{(0,1)})^{k_2}=\gamma_{\bar k}$.

Reordering the contributions, as we have done for the Klein bottle, one can 
obtain:
\begin{itemize}
\item[a)] $N$ odd, $M$ odd:\\
From $D3$-branes:
\beq
\cM_3 = -2\sum_{\bar k=(1,1)}^{(N,M)}\frac{\delta_{\bar k,3}}
                   {8\tilde{s}_1\tilde{s}_2\tilde{s}_3} 
        [32 s_1s_2s_3]  [8 \tilde{s}_1\tilde{s}_2\tilde{s}_3
       \Tr{\gamma_{2\bar k,3}}].
\eeq
\\
From $D7_i$-branes:
\beq
\cM_{7_i} = 2 \sum_{\bar k=(1,1)}^{(N,M)} \frac{\delta_{\bar k,{7_i}}}
                      {8 \tilde{s}_1 \tilde{s}_2 \tilde{s}_3}  
              [32 s_1s_2s_3]  [2 \tilde{s}_i \Tr{\gamma_{2\bar k,{7_i}}}].
\eeq

\item[b)] $N$ odd, $M$ even:\\
From $D3$-branes:
\beq
\cM_3   =  - 2 \sum_{\bar k=(1,1)}^{(N,M/2)} \frac{\delta_{\bar k,3}}
                        {8 \tilde{s}_1 \tilde{s}_2 
              \tilde{s}_3}  [32 (s_1s_2s_3 - \tilde\eps_1 s_1 c_2 c_3)]  
          [8 \tilde{s}_1 \tilde{s}_2 \tilde{s}_3 \Tr{\gamma_{2\bar k,3}}]. 
\eeq
\\
From $D7_i$-branes: 
\beq
\cM_{7_i}   =  2  \sum_{\bar k=(1,1)}^{(N,M/2)} \frac{\delta_{\bar k,{7_i}}}
                            {8 \tilde{s}_1 \tilde{s}_2 \tilde{s}_3}  
                            [32 (s_1s_2s_3 - \tilde\eps_1 s_1 c_2 c_3) ] 
                           [2 \tilde{s}_i \Tr{\gamma_{2\bar k,{7_i}}}].
\eeq

This factorization is necessary to match the cylinder and the Klein bottle 
contributions. But to obtain it, we had to impose some restrictions 
on the `vector structures' $\mu_i$ (introduced in (\ref{def_mu})) and 
the signs $\tilde\eps_i$:
\beq  \label{vec_struc1}
{\mu_1}^3 = {\mu_1}^{7_1} = -{\mu_1}^{7_2} = -{\mu_1}^{7_3} 
= \tilde\eps_1.
\eeq
Here we took $\delta_{\bar k}=1\ \forall\,\bar k$ if $\gamma_\Omega$ is
symmetric and $\delta_{\bar k}=-1\ \forall\,\bar k$ if $\gamma_\Omega$ is
antisymmetric, as in (\ref{orien_fs}).

\item[c)] $N$ even, $M$ even:\\
From $D3$-branes:
\beq
\cM_3 = - 2 \sum_{\bar k=(1,1)}^{(N/2,M/2)} 
                    \frac{\eps^{-k_1k_2}\delta_{\bar k,3}}
                         {8\tilde{s}_1\tilde{s}_2\tilde{s}_3}
       [32 (s_1s_2s_3 - \sum_{i\neq j\neq k}\tilde\eps_i \beta_i s_i c_j c_k)]
       [8 \tilde{s}_1 \tilde{s}_2 \tilde{s}_3 \Tr{\gamma_{2\bar k,3}}].
\eeq
\\
From $D7_i$-branes:
\beq
\cM_{7_i}   =  2  \sum_{\bar k=(1,1)}^{(N/2,M/2)} 
                          \frac{\eps^{-k_1k_2}\delta_{\bar k,{7_i}}}
                               {8\tilde{s}_1\tilde{s}_2\tilde{s}_3}  
       [32 (s_1s_2s_3 - \sum_{i\neq j\neq k}\tilde\eps_i \beta_i s_i c_j c_k)]
                    [2 \tilde{s}_i \Tr{\gamma_{2\bar k,{7_i}}}].
\eeq

Again, this factorization is necessary to match the cylinder and the Klein 
bottle contributions, but it is only possible if we impose some restrictions 
on the `vector structures' $\mu_i$ and the signs $\tilde\eps_i$:
\beqa
{\mu_1}^3 = {\mu_1}^{7_1} = -{\mu_1}^{7_2} = -{\mu_1}^{7_3} 
 &= &\tilde\eps_1, \nonumber\\
{\mu_3}^3 = -{\mu_3}^{7_1} = -{\mu_3}^{7_2} = {\mu_3}^{7_3} 
 &= &\tilde\eps_3,\\
\tilde\eps_1\tilde\eps_2\tilde\eps_3 &= &\eps^{-MN/4}. \nonumber
\eeqa

One further condition is needed, to be able to write the total tadpole
contribution $\cC+\cM+\cK$ as a square\footnote{This condition is necessary
if one uses the standard $\Omega$-projection of \cite{gp}. Using, however,
the alternative projection of \cite{dp}, this condition can be relaxed.}:
\be  \label{delta_cond}
  \delta_{\bar k,3}
  =-\delta_{\bar k,7_1}=-\delta_{\bar k,7_2}=-\delta_{\bar k,7_3}
  \equiv\delta_{\bar k}.
\ee
If the above conditions are satisfied, the tadpole conditions can be written
in the form that appears in the main text, eqs.\ (\ref{tad_NModd})%
--(\ref{tadNMeven}).

\end{itemize}

\section{Shift formalism and discrete torsion} 
\label{app_shift}

Let us sketch a general procedure for obtaining the open string spectrum 
of an arbitrary type IIB orbifold or orientifold with discrete torsion. 
The method is based on the relation between the shift formalism and the 
one using matrices.

The open string spectrum of a general type IIB orbifold with or without
discrete torsion is determined by the matrices $\lambda$ that satisfy
\be  \label{orbspec}
\gamma_g\lambda\gamma_g^{-1}=r(g)\lambda,
\ee
where $\gamma_g$ represents the action of the orbifold group $\Gamma$ on
the Chan-Paton indices of the open string and $r(g)$ the action of $\Gamma$
on the oscillator state of the open string. For the gauge degrees of freedom,
$\lambda^{(0)}$, one has $r(g)=1$, whereas $r(g)=e^{2\pi iv_i}$ for the
matter degrees of freedom associated to the $i$-th complex plane, 
$\lambda^{(i)}$, if $g$ acts as $z_i\to e^{2\pi iv_i}z_i$ on the $i$-th
complex coordinate. A straightforward but cumbersome method to get the
the spectrum, consists in constructing explicitely the matrices 
$\lambda^{(0)}$, $\lambda^{(i)}$ that solve (\ref{orbspec}).
The $\lambda^{(0)}$ transform in the adjoint representation of the gauge
group $G$. Knowing $G$, one finds the matter representations by looking at
the transformation properties of the $\lambda^{(i)}$. 

In \cite{afiv} it has been shown that this is equivalent to a shift 
formalism which is very similar to the one that is known from heterotic
orbifolds. If all the matrices $\gamma_g$ commute, one can diagonalize
them simultaneously by choosing the Cartan-Weyl basis:
\be  \label{gamma_CW}
\gamma_g=e^{-2\pi iV_g\cdot H},
\ee
where $H=(H_1,H_2,\ldots,H_{\rank(\tilde G)})$ is a vector containing the 
Cartan generators of the gauge group $\tilde G$ (before the orbifold 
projection) and $V_g=(V_1,V_2,\ldots,V_{\rank(\tilde G)})$ is the shift 
vector that represents the action of $g\in\Gamma$ on the root lattice of 
$\tilde G$. One can show \cite{afiv} that in the Cartan-Weyl basis the 
identity
\be  \label{gamEgam}
\gamma_g E_a\gamma_g^{-1}=e^{-2\pi i\rho^a\cdot V_g}E_a,\qquad
a=1,\ldots,\dim(\tilde G)-\rank(\tilde G),
\ee
holds for all charged generators $E_a$ of $\tilde G$, if $\rho^a$ 
is the root vector associated to $E_a$, i.e.\ 
$[H_I,E_a]=\rho^a_IE_a$. The condition (\ref{orbspec}) now reads
\bea  \label{orbspec_CW}
\rho^a\cdot V_g &= &0\ {\rm mod}\ \IZ\qquad \hbox{for gauge fields},\\
\rho^a\cdot V_g &= &v_i\ {\rm mod}\ \IZ\qquad \hbox{for matter fields
                                    from the $i$-th complex plane}.
\eea
In this representation, it is very easy to find the gauge group 
$G\subset\tilde G$ that is preserved after the orbifold projection 
and the representations of the matter fields.

If $\Gamma=\IZ_N\times\IZ_M$ and the projective representation $\gamma$ has
discrete torsion, then $\gamma_{g_1}\gamma_{g_2}\neq\gamma_{g_2}\gamma_{g_1}$,
where $g_1,g_2$ are the generators of $\IZ_N,\IZ_M$.
The above method must be modified because $\gamma_{g_1}$ and $\gamma_{g_2}$
cannot be simultaneously diagonalized. Following the idea of \cite{afiu,afiuv},
we diagonalize $\gamma_{g_1}$ and represent $g_2$ by a permutation acting
on the entries of the root vectors.

For a \ZNM\ orbifold with discrete torsion $\eps$, the matrices 
$\gamma_{g_1}$, $\gamma_{g_2}$ are of the following form 
(see appendix \ref{app_proj}):
\bea  \label{gam}
\gamma_{g_1} &= &\bigoplus_{k}\,(\omega_N^k\gamma_{g_1}^{\rm irrep})
                                \otimes\one_{n_{k}},\qquad
\gamma_{g_2}\ =\ \bigoplus_{l}\,(\omega_M^l\gamma_{g_2}^{\rm irrep})
                                \otimes\one_{n_{l}},\\
{\rm with} &&\gamma_{g_1}^{\rm irrep}\ =\ 
                   \diag(1,\eps^{-1},\eps^{-2},...,\eps^{-(s-1)}),\qquad
              \gamma_{g_2}^{\rm irrep}\ =\ 
                \left(\ba{ccccc}
                  0 & 1 & 0 & \dots &0\\
                  0 & 0 & 1 & \dots & \\
                  \vdots & & \ddots & \ddots \\
                  0 &   & \dots & 0 &1\\
                  1 & 0 & \dots &  &0
                  \ea\right),   \nonumber\\
          &&\omega_p=e^{2\pi i/p},\qquad k=0,\ldots,{N\over s}-1,\quad
                                         l=0,\ldots,{M\over s}-1, \nonumber\\
          &&s \quad\hbox{is the minimal positive integer, such that }\eps^s=1.
            \nonumber
\eea
In the Cartan-Weyl basis, $\gamma_{g_1}$ corresponds to the shift
\bea  \label{shift_V}
V_{g_1} &= &{1\over N}\Bigg(0^{n_0},\left({N\over s}\right)^{n_0},\ldots,
            \left((s-1){N\over s}\right)^{n_0},
            1^{n_1},\left({N\over s}+1\right)^{n_1},\ldots,
            \left((s-1){N\over s}+1\right)^{n_1},\nonumber\\
         &&\qquad\ldots,\left({N\over s}-1\right)^{n_{N/s-1}},
           \left(2{N\over s}-1\right)^{n_{N/s-1}},\ldots,
           (N-1)^{n_{N/s-1}}\Bigg).
\eea 
The matrix $\gamma_{g_2}$ acts as a permutation $\Pi_{g_2}$ on the
root vectors $\rho$. We assume $N\ge M$ and consider first the case 
$s=M$, i.e.\ $\gamma_{g_2}=\gamma_{g_2}^{\rm irrep}\otimes\one_n$, 
where $n=\sum_k n_k$:
\bea  \label{perm_Pi}
\Pi_{g_2}\,: &&(\tilde\rho_1^{(n_0)},\ldots,\tilde\rho_s^{(n_0)},
                \tilde\rho_1^{(n_1)},\ldots,\tilde\rho_s^{(n_1)},\ldots,
                \tilde\rho_1^{(n_{N/s-1})},\ldots,\tilde\rho_s^{(n_{N/s-1})})
             \\
      &&\longrightarrow\ (\tilde\rho_s^{(n_0)},\tilde\rho_1^{(n_0)},\ldots,
                                   \tilde\rho_{s-1}^{(n_0)},
           \tilde\rho_s^{(n_1)},\tilde\rho_1^{(n_1)},\ldots,
                                   \tilde\rho_{s-1}^{(n_1)},\ldots,
           \tilde\rho_s^{(n_{N/s-1})},\ldots,\tilde\rho_{s-1}^{(n_{N/s-1})}),
            \nonumber\\
{\rm where}   &&\tilde\rho_i^{(n_k)}=(\rho_{s(n_0+\ldots+n_{k-1})+(i-1)n_k+1},
                                \ldots,\rho_{s(n_0+\ldots+n_{k-1})+in_k}).
               \nonumber
\eea
The gauge group $G$ of the orbifold model is determined by finding all the
root vectors $\rho^a$ of $U(sn)$ (or linear combinations of these) that
satisfy
\be  \label{gauge_cond}
\rho^a\cdot V_{g_1}=0\ {\rm mod}\ \IZ,\qquad \Pi_{g_2}(\rho^a)=\rho^a.
\ee
If we choose twist vectors $v={1\over N}(1,-1,0)$ resp.\
$w={1\over M}(0,1,-1)$ for the action of $g_1$ resp.\ $g_2$ on 
$\IC^3$, then the matter representations are obtained from the
root vectors $\rho^a$ of $U(sn)$ (or linear combinations of these) that
satisfy
\be  \label{matter_cond}
\ba{lll}
\rho^a\cdot V_{g_1}={1\over N}\ {\rm mod}\ \IZ,\quad 
    &\Pi_{g_2}(\rho^a)=\rho^a \quad 
    &\hbox{($1^{\rm st}$ complex plane)\quad or}\\ 
\rho^a\cdot V_{g_1}=-{1\over N}\ {\rm mod}\ \IZ,\quad 
    &\Pi_{g_2}(\rho^a)=e^{2\pi i/M}\rho^a \quad 
    &\hbox{($2^{\rm nd}$ complex plane)\quad or}\\
\rho^a\cdot V_{g_1}=0\ {\rm mod}\ \IZ,\quad 
    &\Pi_{g_2}(\rho^a)=e^{-2\pi i/M}\rho^a \quad 
    &\hbox{($3^{\rm rd}$ complex plane)}.
\ea
\ee

Consider, for example, $\Gamma=\IZ_N\times\IZ_N$ with discrete torsion
$\eps=\omega_N$, i.e.\ $s=N$ (these models were discussed in \cite{df}). 
In this case, one has
\bea  \label{shifts_NN}
&&V_{g_1}\ =\ {1\over N}\left(0^{n_0},1^{n_0},\ldots,(N-1)^{n_0}\right), \\
&&\Pi_{g_2}\,:\ (\tilde\rho_1^{(n_0)},\ldots,\tilde\rho_N^{(n_0)})\ 
         \longrightarrow\ (\tilde\rho_N^{(n_0)},\tilde\rho_1^{(n_0)},\ldots,
                                   \tilde\rho_{N-1}^{(n_0)}). \nonumber
\eea
The roots of $U(Nn_0)$ are of the form $\rho^a=(\underline{+,-,0^{Nn_0-2}})$,
where underlining means that all permutations have to be considered. It is
easy to see that there are $n_0(n_0-1)$ linear combinations of roots $\rho_a$
that satisfy (\ref{gauge_cond}):
$$ (\underline{+,-,0^{n_0-2}},0^{(N-1)n_0})+
   (0^{n_0},\underline{+,-,0^{n_0-2}},0^{(N-2)n_0})+\ldots+
   (0^{(N-1)n_0},\underline{+,-,0^{n_0-2}}).$$
Together with the $n_0$ Cartan generators they form the gauge group 
$G=U(n_0)$. Similarly, one finds $3\,n_0^2-n_0$ linear combinations of roots 
$\rho_a$ (and $n_0$ Cartan generators) that satisfy (\ref{matter_cond}):
\begin{eqnarray*}
1^{\rm st}\ {\rm plane}: 
    &&(\underline{-,0^{n_0-1}},\underline{+,0^{n_0-1}},0^{(N-2)n_0})+
         (0^{n_0},\underline{-,0^{n_0-1}},\underline{+,0^{n_0-1}},
          0^{(N-3)n_0})\\
    &&\qquad+\ldots+(\underline{+,0^{n_0-1}},0^{(N-2)n_0},
                     \underline{-,0^{n_0-1}}),\\
2^{\rm nd}\ {\rm plane}:
    &&(\underline{+,0^{n_0-1}},\underline{-,0^{n_0-1}},0^{(N-2)n_0})+
      \alpha^{-1}\,(0^{n_0},\underline{+,0^{n_0-1}},\underline{-,0^{n_0-1}},
          0^{(N-3)n_0})\\
    &&\qquad+\ldots+\alpha^{-(N-1)}\,(\underline{-,0^{n_0-1}},0^{(N-2)n_0},
                                      \underline{+,0^{n_0-1}}),\\
3^{\rm rd}\ {\rm plane}:
    &&(\underline{+,-,0^{n_0-2}},0^{(N-1)n_0})+
      \alpha\,(0^{n_0},\underline{+,-,0^{n_0-2}},0^{(N-2)n_0})\\
    &&\qquad+\ldots+\alpha^{N-1}\,(0^{(N-1)n_0},\underline{+,-,0^{n_0-2}}),
\end{eqnarray*}
where $\alpha=e^{2\pi i/N}$. These are three matter fields in the adjoint
representation of $U(n_0)$.

As a second example, consider $\Gamma=\IZ_N\times\IZ_2$ with discrete torsion
$\eps=-1$, i.e.\ $s=2$. In this case, one has
\bea  \label{shifts_N2}
&&\hspace{-5mm}V_{g_1}\ =\ {1\over N}\left(0^{n_0},
            \left({N\over2}\right)^{n_0},1^{n_1},
            \left({N\over2}+1\right)^{n_1},\ldots,
            \left({N\over2}-1\right)^{n_{N/2-1}},(N-1)^{n_{N/2-1}}\right), 
            \nonumber\\
&&\hspace{-5mm}\Pi_{g_2}\,:\ (\tilde\rho_1^{(n_0)},\tilde\rho_2^{(n_0)},
                 \tilde\rho_1^{(n_1)},\tilde\rho_2^{(n_1)},\ldots,
                 \tilde\rho_1^{(n_{N/2-1})},\tilde\rho_2^{(n_{N/2-1})})\\ 
         &&\hspace{10mm}\longrightarrow\ (\tilde\rho_2^{(n_0)},
                 \tilde\rho_1^{(n_0)},
                 \tilde\rho_2^{(n_1)},\tilde\rho_1^{(n_1)},\ldots,
                 \tilde\rho_2^{(n_{N/2-1})},\tilde\rho_1^{(n_{N/2-1})}). 
                 \nonumber 
\eea
There are $n_0(n_0-1)+n_1(n_1-1)+\ldots+n_{N/2-1}(n_{N/2-1}-1)$
linear combinations of roots of $U(2n)$ that satisfy (\ref{gauge_cond}).
Together with the $n=\sum_kn_k$ Cartan generators, they form the gauge
group $G=U(n_0)\times U(n_1)\times\cdots\times U(n_{N/2-1})$. Similarly,
one finds the following matter representations from the 3 complex planes:
\bea
1^{\rm st}\ {\rm plane}: && (\antifund,\fund,\one,\ldots,\one) 
      +(\one,\antifund,\fund,\one,\ldots,\one)+\cdots
      +(\one,\ldots,\one,\antifund,\fund) 
      +(\fund,\one,\ldots,\one,\antifund) \nonumber\\
2^{\rm nd}\ {\rm plane}: && (\fund,\antifund,\one,\ldots,\one) 
      +(\one,\fund,\antifund,\one,\ldots,\one)+\cdots
      +(\one,\ldots,\one,\fund,\antifund) 
      +(\antifund,\one,\ldots,\one,\fund) \nonumber\\
3^{\rm rd}\ {\rm plane}: && (adj,\one,\ldots,\one)+(\one,adj,\one,\ldots,\one)
                           +\ldots+(\one,\ldots,\one,adj) \nonumber
\eea

It is instructive to see in a simple special case that the permutation
$\Pi_{g_2}$ in the shift formalism indeed corresponds to the action of
$\gamma_{g_2}$ in the formalism using matrices. Let us take 
$\Gamma=\IZ_4\times\IZ_2$. According to (\ref{gam}) the matrices
$\gamma_{g_1}$, $\gamma_{g_2}$ are given by
\be  \label{Z4Z2matrix}
\gamma_{g_1} \ =\ \left(\begin{array}{cccc}
                  \one_{n_0}& 0 & 0 & 0 \\
                  0 & -\one_{n_0} & 0 & 0 \\
                  0 & 0 & i\one_{n_1} & 0 \\
                  0 & 0 & 0 & -i\one_{n_1}
                  \end{array}\right),\quad
\gamma_{g_2} \ =\ \left(\begin{array}{cccc}
                  0 & \one_{n_0} & 0 & 0 \\
                  \one_{n_0} & 0 & 0 & 0 \\
                  0 & 0 & 0 & \one_{n_1} \\
                  0 & 0 & \one_{n_1} & 0
                  \end{array}\right)
\ee
As $\gamma_{g_1}$ is diagonal, according to (\ref{gamma_CW}), it
corresponds to a shift $V_{g_1}=\frac14(0^{n_0},2^{n_0},1^{n_1},3^{n_1})$.
The action of $\gamma_{g_2}E_a\gamma_{g_2}^{-1}$ on the matrix $E_a$ is
such that it permutes the associated roots $\rho^a$ as
$(\tilde\rho^{(n_0)}_1,\tilde\rho^{(n_0)}_2,\tilde\rho^{(n_1)}_1,
\tilde\rho^{(n_1)}_2)\ \to\ (\tilde\rho^{(n_0)}_2,\tilde\rho^{(n_0)}_1,
\tilde\rho^{(n_1)}_2,\tilde\rho^{(n_1)}_1)$. This coincides with the
prescription given above. More precisely, $\gamma_{g_2}E_a\gamma_{g_2}^{-1}
=\alpha E_a^\prime$, where $E_a^\prime$ is the matrix corresponding to the
permuted root vector. The phase $\alpha=e^{-m\pi i/M}$, $m=0,1,2,3$, is
related to an ambiguity in choosing a specific basis for the matrices $E_a$.
It does not affect the orbifold spectrum and can therefore be ignored.
However, in the orientifold case, it will be important to take this phase 
into account.

For a general \ZNM\ orbifold with discrete torsion $\eps$, such that $s<M$,
one has to take $M/s$ copies of the shift vector (\ref{shift_V}),
with entries labelled by
$$(n_{0,0},n_{1,0},\ldots,n_{N/s-1,0}|n_{0,1},n_{1,1},\ldots,n_{N/s-1,1}|
   \cdots|n_{0,M/s-1},n_{1,M/s-1},\ldots,n_{N/s-1,M/s-1}).$$
The permutation $\Pi_{g_2}$ acts on each copy identically, as in 
(\ref{perm_Pi}). To represent the action of the phases $\omega_M^l$ that
appear in (\ref{gam}) for $s<M$, one has to associate a shift vector also
to the element $g_2$:
\be  \label{shift_Vg2}
V_{g_2}\ =\ {1\over M}\left(0^{\tilde n_0},1^{\tilde n_1},\ldots,
            \left({M\over s}-1\right)^{\tilde n_{M/s-1}}\right),\qquad
{\rm with}\quad\tilde n_l=\sum_kn_{k,l}. \nonumber
\ee
To determine the gauge group, one has to find all roots $\rho^a$ of
$U(s\sum_{k,l}n_{k,l})$ that satisfy
\be  \label{gen_gauge_cond}
\rho^a\cdot V_{g_1}=0\ {\rm mod}\ \IZ,\qquad
\rho^a\cdot V_{g_2}=0\ {\rm mod}\ \IZ,\qquad
\Pi_{g_2}(\rho^a)=\rho^a.
\ee
The roots corresponding to the matter representations have to satisfy,
for each complex plane, {\em one} of the following $s$ conditions:
\bea  \label{gen_matter_cond}
&&\rho^a\cdot V_{g_1}=v_i\ {\rm mod}\ \IZ,\qquad 
  \rho^a\cdot V_{g_2}=w_i-{r\over s}\ {\rm mod}\ \IZ,\qquad 
  \Pi_{g_2}(\rho^a)=e^{2\pi i r/s}\rho^a,\nonumber\\
&&i=1,2,3,\qquad r=0,\ldots,s-1, \\
&&v_1={1\over N},\ v_2=-{1\over N},\ v_3=0,\ 
  w_1=0,\ w_2={1\over M},\ w_3=-{1\over M}.\nonumber
\eea

As an example, let us compute the spectrum of the $\IZ_6\times\IZ_6$
orbifold with discrete torsion $\eps=e^{2\pi i/3}$, i.e.\ $s=3$.
The shift vectors and permutation are given by
\bea  \label{shift_66}
V_{g_1} &= &{1\over6}\Big(0^{n_{0,0}},2^{n_{0,0}},4^{n_{0,0}},1^{n_{1,0}},
                 3^{n_{1,0}},5^{n_{1,0}}|0^{n_{0,1}},2^{n_{0,1}},4^{n_{0,1}},
                 1^{n_{1,1}},3^{n_{1,1}},5^{n_{1,1}}\Big), \nonumber\\
V_{g_2} &= &{1\over6}\Big(0^{\tilde n_0},1^{\tilde n_1}\Big),\\
\Pi_{g_2}: &&\hspace{-5mm} (\tilde\rho_1^{n_{0,0}},\tilde\rho_2^{n_{0,0}},
               \tilde\rho_3^{n_{0,0}},\ldots,\tilde\rho_1^{n_{1,1}},
               \tilde\rho_2^{n_{1,1}},\tilde\rho_3^{n_{1,1}})
            \ \to\ (\tilde\rho_3^{n_{0,0}},\tilde\rho_1^{n_{0,0}},
               \tilde\rho_2^{n_{0,0}},\ldots,\tilde\rho_3^{n_{1,1}},
               \tilde\rho_1^{n_{1,1}},\tilde\rho_2^{n_{1,1}}).\nonumber
\eea
There are $n_{0,0}(n_{0,0}-1)$ linear combinations of roots,
$$(\underline{+,-,0^{n_{0,0}-2}},0,\ldots,0)+
  (0^{n_{0,0}},\underline{+,-,0^{n_{0,0}-2}},0,\ldots,0)+
  (0^{2n_{0,0}},\underline{+,-,0^{n_{0,0}-2}},0,\ldots,0),$$
that satisfy (\ref{gen_gauge_cond}), and similarly for $n_{1,0}$,
$n_{0,1}$, $n_{1,1}$. Together with the Cartan generators, they form the
gauge group $U(n_{0,0})\times U(n_{1,0})\times U(n_{0,1})\times U(n_{1,1})$.
Concerning the matter fields from the first plane, only $r=0$ is possible in
(\ref{gen_matter_cond}). This gives the following representations:
$$(\fund,\antifund,\one,\one)+(\antifund,\fund,\one,\one)+
  (\one,\one,\fund,\antifund)+(\one,\one,\antifund,\fund).$$
In the second plane, $r=0$ and $r=1$ are possible, yielding
$$r=0:\ (\one,\antifund,\fund,\one)+(\antifund,\one,\one,\fund),\qquad
  r=1:\ (\one,\fund,\antifund,\one)+(\fund,\one,\one,\antifund).$$
In the third plane, $r=0$ and $r=2$ are possible, yielding
$$r=0:\ (\fund,\one,\antifund,\one)+(\one,\fund,\one,\antifund),\qquad
  r=2:\ (\antifund,\one,\fund,\one)+(\one,\antifund,\one,\fund).$$

The open string spectrum of a general type IIB orientifold with or without
discrete torsion is determined by the matrices $\lambda$ that satisfy
(\ref{orbspec}) and, in addition,
\be  \label{orispec}
\gamma_\Omega\lambda^\top\gamma_\Omega^{-1}=r(\Omega)\lambda,
\ee
where $\gamma_\Omega$ represents the action of the world-sheet parity
$\Omega$ on the Chan-Paton indices of the open string and $r(\Omega)$ the 
action of $\Omega$ on the oscillator state of the open string.
In complete analogy to the orbifold case, the spectrum can equivalently
be obtained using the shift formalism. The additional condition 
(\ref{orispec}) implies that the gauge group $\tilde G$ before the
orbifold projection is $SO(4\sum_{k,l}n_{k,l})$ resp.\ 
$USp(4\sum_{k,l}n_{k,l})$ if $\gamma_\Omega$ is symmetric resp.\
antisymmetric. In the case of orientifolds with discrete torsion, one
has to find all root vectors $\rho^a$ of $\tilde G$ that satisfy
(\ref{gen_gauge_cond}), (\ref{gen_matter_cond}).
As a consequence of $\Omega g\Omega=g$, the matrices $\gamma_g$ now 
form a real projective representation and only discrete torsion $\eps=\pm1$
is possible.

In this case one must also distinguish between orientifolds with four
different boundary conditions on the gauge bundle: $\gamma_{g_1}^N=\pm\one$,
$\gamma_{g_2}^M=\pm\one$. In the shift formalism this translates to
\be
NV_{g_1}=\left\{\ba{ll} (1,\ldots,1) \ {\rm mod} \ \IZ
                                 &{\rm if\ }\gamma_{g_1}^N=\one\\
                  \half(1,\ldots,1) \ {\rm mod} \ \IZ 
                                 &{\rm if\ }\gamma_{g_1}^N=-\one
                 \ea\right.,
\ee
and similarly for $MV_{g_2}$. As mentioned above (below eq.\ 
(\ref{Z4Z2matrix})), there is an ambiguity in the choice of a basis for
the Chan-Paton matrices $\lambda$ which leads to an additional phase in
the action of $g_2$ on $\lambda$. It turns out that the above prescription,
using a shift $V_{g_2}$ of the form (\ref{shift_Vg2}) and requiring the
conditions (\ref{gen_gauge_cond}), (\ref{gen_matter_cond}), corresponds
in the orientifold case to $\gamma_{g_2}^M=-\one$. However, using the shift
$V_{g_1}$ of the form (\ref{shift_V}) for the group element $g_1$, 
corresponds to $\gamma_{g_1}^N=\one$, as expected. If $\gamma_{g_1}^N=-\one$, 
$\gamma_{g_2}^M=\one$, then the general form of the shifts given above, 
eqs.\ (\ref{shift_V}), (\ref{shift_Vg2}), is modified to (using $s=2$ and 
taking $M/2$ copies of (\ref{shift_V}))
\bea  \label{mod_shifts}
V_{g_1} &= &{1\over2N}\Big(1^{n_{0,0}},(N+1)^{n_{0,0}},
            3^{n_{1,0}},(N+3)^{n_{1,0}},\ldots,
            (N-1)^{n_{N/2-1,0}},(2N-1)^{n_{N/2-1,0}}|
          \nonumber\\
 &&\qquad\qquad 1^{n_{0,1}},(N+1)^{n_{0,1}},\ldots|\cdots|\ldots
          (2N-1)^{n_{N/2-1,M/2-1}}\Big),\\   
V_{g_2} &= &{1\over2M}\Big(1^{\tilde n_0},3^{\tilde n_1},\ldots,
             (M-1)^{\tilde n_{M/2-1}}\Big),\qquad
{\rm with}\quad\tilde n_l=\sum_kn_{k,l}. \nonumber
\eea
The permutation $\Pi_{g_2}$ is not modified. The gauge group is determined
by the roots $\rho^a$ that satisfy {\em one} of the two conditions
\be  \label{gauge_orien}
\rho^a\cdot V_{g_1}=0\ {\rm mod}\ \IZ,\qquad
\rho^a\cdot V_{g_2}={r\over2}\ {\rm mod}\ \IZ,\qquad
\Pi_{g_2}(\rho^a)=(-1)^r\,\rho^a,\qquad r=0,1.
\ee
The matter fields are obtained from (\ref{gen_matter_cond}) by setting $s=2$.

\end{appendix}

\end{document}